\UseRawInputEncoding
\documentclass[nolinenumbers]{aastex631}

\shorttitle{Cloud variability}
\shortauthors{Cohen et al.}
\begin{document}

\title{Traveling planetary-scale waves cause cloud variability on tidally locked aquaplanets}

\author[0000-0001-5014-4174]{Maureen Cohen}
\affiliation{School of GeoSciences,
The University of Edinburgh \\
Edinburgh, EH9 3FF, UK}
\affiliation{Centre for Exoplanet Science, The University of Edinburgh, UK}

\author[0000-0001-7509-7650]{Massimo A. Bollasina}
\affiliation{School of GeoSciences,
The University of Edinburgh \\
Edinburgh, EH9 3FF, UK}

\author[0000-0001-8832-5288]{Denis E. Sergeev}
\affiliation{Department of Mathematics \\
College of Engineering, Mathematics, and Physical Sciences, University of Exeter \\
Exeter, EX4 4QF, UK}

\author[0000-0002-1487-0969]{Paul I. Palmer}
\affiliation{School of GeoSciences,
The University of Edinburgh \\
Edinburgh, EH9 3FF, UK}
\affiliation{Centre for Exoplanet Science, The University of Edinburgh, UK}

\author[0000-0001-6707-4563]{Nathan J. Mayne}
\affiliation{Department of Astrophysics \\
College of Engineering, Mathematics, and Physical Sciences, University of Exeter \\
Exeter, EX4 4QF, UK}

\correspondingauthor{Maureen Cohen}
\email{s1144983@ed.ac.uk}


\begin{abstract}

Cloud cover at the planetary limb of water-rich Earth-like planets is likely to weaken chemical signatures in transmission spectra, impeding attempts to characterize these atmospheres. However, based on observations of Earth and solar system worlds, exoplanets with atmospheres should have both short-term weather and long-term climate variability, implying that cloud cover may be less during some observing periods. We identify and describe a mechanism driving periodic clear sky events at the terminators in simulations of tidally locked Earth-like planets. A feedback between dayside cloud radiative effects, incoming stellar radiation and heating, and the dynamical state of the atmosphere, especially the zonal wavenumber-1 Rossby wave identified in past work on tidally locked planets, leads to oscillations in Rossby wave phase speeds and in the position of Rossby gyres and results in advection of clouds to or away from the planet's eastern terminator. We study this oscillation in simulations of Proxima Centauri b, TRAPPIST 1-e, and rapidly rotating versions of these worlds located at the inner edge of their stars' habitable zones. We simulate time series of the transit depths of the 1.4 $\mu$m water feature and 2.7 $\mu$m carbon dioxide feature. The impact of atmospheric variability on the transmission spectra is sensitive to the structure of the dayside cloud cover and the location of the Rossby gyres, but none of our simulations have variability significant enough to be detectable with current methods.

\end{abstract}



\section{Introduction} \label{sec:intro}
The capabilities of the James Webb Space Telescope (JWST) have raised the prospect of characterizing the atmospheres of transiting exoplanets through transmission spectroscopy \citep{molliere_observing_2017,greene_characterizing_2016,beichman_observations_2014}. Particular interest has focused on the characterization of rocky and temperate planets orbiting at distances from their host stars that would allow liquid water to exist on their surfaces \citep{gialluca_characterizing_2021,morley_observing_2017}. A number of terrestrial planets have been found in this range of orbital distances, known as the habitable zone \citep{kasting_habitable_1993}, including the non-transiting Proxima Centauri b \citep{anglada-escude_terrestrial_2016} in orbit around the closest star to Earth, Proxima Centauri, and three transiting planets in orbit around the star TRAPPIST-1 \citep{gillon_seven_2017}. These planets are thought to be tidally locked to their host stars as a result of their close-in orbits \citep{barnes_tidal_2017}, and indeed tidally locked planets around M-dwarf stars may be the most common type of potentially habitable planet \citep{dressing_occurrence_2015, kopparapu_revised_2013}.

A challenge for transmission spectroscopy of transiting exoplanets is the presence of clouds, which mute spectroscopic features by scattering light isotropically at the level of the cloud deck \citep{barstow_curse_2021, helling_exoplanet_2019}. Clouds are believed to exist on multiple known exoplanets \citep{burningham_cloud_2021, helling_cloud_2021, kreidberg_clouds_2014}. Modelling studies of the impact of clouds on the transmission spectra of water-rich rocky planets have indicated that in most cases it would take anywhere from ten to hundreds of transits to detect atmospheric absorption features using the JWST \citep{komacek_clouds_2020, suissa_dim_2020, fauchez_impact_2019}. Available observations of the planets in the TRAPPIST-1 system have ruled out hydrogen-rich primordial atmospheres for these planets \citep{garcia_hstwfc3_2022, moran_limits_2018, de_wit_combined_2016}, but are unable to break the degeneracy between a cloud- or aerosol-heavy atmosphere, a high molecular mean weight atmosphere, or the absence of an atmosphere, although the JWST may be able to do so in future \citep{lustig-yaeger_detectability_2019}. Some work has offered brighter prospects of the detection of water vapor on arid (icy) planets \citep{ding_prospects_2022} and found that stratospheric (as opposed to tropospheric) clouds would not necessarily affect observations by the JWST \citep{doshi_stratospheric_2022}. As water-rich planets are expected to form substantial cloud decks, this limitation is a significant obstacle to the detection of atmospheric chemistry and potential biosignatures on water-rich habitable worlds.

One possible avenue for characterizing water-rich planets is temporal variability in cloud cover. Studies of exoplanet variability are extremely limited so far, but variable wind speeds may have been detected on KELT-9b \citep{asnodkar_variable_2022} and variation in the offset of the peak of the phase curve of HAT-P-7b was reported by \cite{armstrong_variability_2016} and later disputed by \cite{lally_reassessing_2022}. Some theoretical \citep{welbanks_atmospheric_2022, powell_transit_2019,line_influence_2016} and observational \citep{mikal-evans_diurnal_2022, ehrenreich_nightside_2020} studies have found that it may be possible to detect spatial variability in cloud cover at the planetary terminators of large exoplanets. In a one-dimensional model, \cite{tan_atmospheric_2019} found that cloud radiative feedback can drive atmospheric variability on brown dwarfs and giant planets.\cite{hochman_greater_2022} used a dynamical systems approach to show that the climate of a tidally locked rocky planet was overall more sensitive to changes in basic parameters (CO$_2$ partial pressure in their study) than that of Earth, noting that tidally locked M-dwarf planets may have climate variability similar to Earth's seasons even with zero obliquity and eccentricity. Of most relevance, \cite{song_asymmetry_2021} \cite{fauchez_trappist-1_2022}, and \cite{may_water_2021} simulated the effect of cloud variability on transmission spectra and atmospheric retrievals of TRAPPIST-1e. \cite{song_asymmetry_2021} found both spatial asymmetry in transit depths when comparing the eastern and western terminators and temporal variability in the transmission spectra. In their study using the ExoCAM general circulation model, the authors found no periodicity in the time series of transit depths. In \cite{may_water_2021}, general circulation model simulations also performed with ExoCAM likewise exhibited cloud cover variability at the planetary limb. The authors combined ten synthetic spectra randomly chosen from a time series of 365 days of the planet's climate and used the resulting composite spectrum to retrieve atmospheric chemical abundances, finding that this did not result in a difference compared to the use of non-variable spectra. However, \cite{may_water_2021} did not study the cause of the cloud variability in their simulations or look for periodicities. An understanding of the physics of cloud and climate variability is necessary to confirm that this variability is not noise and to explain why different models predict vastly different degrees of variability.

In this work, we describe a dynamical mechanism driving cloud and climate variability in the atmospheres of moist tidally locked terrestrial exoplanets and investigate its impact on time series of transmission spectra. In Section \ref{sec:methods}, we describe our general circulation model, simulation parameter space, and radiative transfer scheme for simulating transmission spectra. In Section \ref{sec:results}, we outline a feedback loop between cloud radiative effects, incoming stellar radiation, and the dynamical state of the atmosphere that causes back-and-forth propagation or shifting of planetary-scale (Rossby) waves and regular variations in cloud cover at the planetary limb. We further discuss the interaction between the propagating Rossby gyres and the dayside cloud structure and simulate time series for the water absorption feature at 1.4 $\mu$m and the carbon dioxide feature at 2.7 $\mu$m. Our results support the findings of \cite{song_asymmetry_2021}, \cite{may_water_2021} and \cite{fauchez_trappist-1_2022} that cloud variability is unlikely to affect JWST observations, except in specific cases where the cloud structure and wave propagation may interact in a fortuitous way. In Section \ref{sec:discussion}, we discuss our results in the context of previous work on Rossby wave structures on tidally locked planets, as well as implications of dynamical variability for the planetary climate and for observational practices. We conclude in Section \ref{sec:conclusion}.

\section{Methods} \label{sec:methods}

\subsection{Model description} \label{subsec:model}
Our simulations are based on the Global Atmosphere 7.0 (GA7) configuration of the Met Office Unified Model (UM). Idealized versions of the UM have previously been used to simulate hot Jupiters \citep{christie_impact_2021, mayne_results_2017, mayne_unified_2014,zamyatina_observability_2023} and terrestrial planets \citep{braam_lightning-induced_2022, sergeev_bistability_2022, eager-nash_implications_2020, boutle_exploring_2017}. The model uses the ENDGame (Even Newer Dynamics for General atmospheric modelling of the environment) dynamical core to solve the non-hydrostatic, fully compressible, deep-atmosphere Navier-Stokes equations \citep{wood_inherently_2014}. GA7 contains parameterizations for sub-grid scale turbulence, convection, non-orographic gravity wave drag, boundary layer processes, precipitation, and clouds. Radiative transfer is simulated using the SOCRATES (Suite Of Community RAdiative Transfer codes based on Edwards and Slingo) community radiative transfer code. All simulations are run at a resolution of $2^\circ$ latitude by $2.5^\circ$ longitude. The substellar point is defined to be at $0^{\circ}$ longitude and latitude, while the antistellar point is located at $180^{\circ}$ longitude and the eastern and western terminators at $90^{\circ}$E and $90^{\circ}$W, respectively. ``Days'' refers to Earth days throughout this work. 

The UM has a fully prognostic cloud scheme, the Prognostic Cloud fraction and Prognostic Condensate (PC2) scheme \citep{wilson_pc2_2008}. The scheme has three prognostic cloud fractions (liquid, ice, and mixed-phase), as well as water vapor and liquid and frozen condensate. These prognostic variables are updated in increments by processes in the model, including advection, convection, and precipitation. The column cloud fraction is determined by exponential random overlap. The moist atmosphere configuration includes water vapor with evaporation and precipitation and an otherwise 100 \% nitrogen atmosphere with fixed trace CO$_2$. In the TRAPPIST-1 Habitable Atmosphere Intercomparison (THAI) \citep{sergeev_trappist-1_2022}, the UM's cloud scheme produced a mean cloud fraction in the middle of the comparison (60 \%), compared to the extremes of the Laboratoire de M\'et\'eorologie Dynamique - Generic model (LMD-G, at 28 \%) and the Resolving Orbital and Climate Keys of Earth and Extraterrestrial Environments with Dynamics model (ROCKE-3D, at 77 \%).

\begin{table}
\hspace*{-2cm} 
\begin{tabular}{llllll}
\hline
Parameter & Control ProxB & Warm ProxB & Control TRAP-1e & Warm TRAP-1e & Dry TRAP-1e  \\
\hline \hline
Semi-major axis (AU) & 0.0485 & 0.0423 & 0.029 & 0.025 & 0.029 \\
Stellar irradiance (W m$^{-2}$) & 881.7 & 1100.2 & 837.7 & 1392.9 & 837.7 \\
Orbital period (Earth days) & 11.2 & 9.2 & 6.1 & 4.2 & 6.1 \\
Rotation speed (rad s$^{-1}$) &  6.501$\times$10$^{-6}$ & 7.933$\times$10$^{-6}$ & 1.192$\times$10$^{-5}$ & 1.746$\times$10$^{-5}$ & 1.192$\times$10$^{-5}$ \\
Eccentricity ($\cdot$) & 0 & 0 & 0 & 0 & 0    \\
Obliquity ($\cdot$) & 0 & 0 & 0 & 0 & 0   \\
Radius (km) & 7160 & 7160 & 5797 & 5797 & 5797   \\
Acceleration due to gravity (m s$^{-2}$) & 10.9 & 10.9 & 9.1 & 9.1 & 9.1 \\
CO$_2$ (ppmv) & 378 & 378 & 400 & 400 & 400 \\
Number of levels ($\cdot$) & 60 & 60 & 39 & 39 & 39 \\
Model top (km) & 85 & 85 & 80 & 80 & 80 \\
\hline
\end{tabular}
\caption{Model parameters for all simulations}
\label{tab:parameters}
\end{table}

\subsection{Simulation parameters} \label{subsec:sims}
We performed five simulations: 
\begin{enumerate}
    \item A ``control'' moist Proxima Centauri b with planetary and orbital parameters as described in \cite{anglada-escude_terrestrial_2016} (Control ProxB)
   \item A ``warm'' moist Proxima Centauri b with planetary and orbital parameters corresponding to the inner edge of Proxima Centauri's habitable zone (Warm ProxB)
    \item A ``control'' moist TRAPPIST-1e with planetary and orbital parameters as described in \cite{gillon_seven_2017} (Control TRAP-1e)
    \item A ``warm'' moist TRAPPIST-1e with planetary and orbital parameters corresponding to the inner edge of TRAPPIST-1's habitable zone (Warm TRAP-1e)
    \item A ``dry'' TRAPPIST-1e atmosphere identical to the control case aside from the dry atmosphere (Dry TRAP-1e)
\end{enumerate}

Table \ref{tab:parameters} lists the values of the parameters varied between each simulation. These parameters were chosen to facilitate comparison with previous UM studies of the two planets. The simulation set-up for Proxima Centauri b is based on \cite{boutle_exploring_2017} and \cite{cohen_longitudinally_2022}, using a model top of 85~km with 60 vertical levels, quadratically stretched to give greater resolution near the surface. The planet is simulated with a slab ocean \citep{frierson_gray-radiation_2006} which has a mixing depth of 2.4~m, representing a heat capacity of $10^7$~{J K$^{-1}$ m$^{-2}$}. The stellar spectrum for Proxima Centauri, modelled as a quiescent M-dwarf, was taken from BT-Settl \citep{rajpurohit_effective_2013} with $T_{eff} = 3000$ K, g= $1000 \mathrm{m s^{-2}}$, and metallicity $=0.3$ dex. The Proxima Centauri b simulations were spun up from an equilibrium state of a previous simulation performed using the UM with the same configuration.

For TRAPPIST-1e, we use the simulation parameters of the TRAPPIST-1 Habitable Atmosphere Intercomparison for both the dry and the moist atmosphere cases \citep{fauchez_trappist_2021, turbet_trappist-1_2022,  sergeev_trappist-1_2022, fauchez_trappist-1_2022}. In this instance, the models use 39 vertical levels with a top of 80~km. The planet's surface in the moist case is a slab ocean with a mixing layer of 1~m, representing a heat capacity of $4 \times 10^6$~{J K$^{-1}$ m$^{-2}$}. The spectrum is taken from BT-Settl \citep{rajpurohit_effective_2013} with $T_{eff} = 2600$ K and Fe/H=0. The TRAPPIST-1e simulations were spun up from an initial state of an isothermal (300K) dry atmosphere at rest with zero winds, following the THAI protocol \citep{fauchez_trappist-1_2020}. Unlike in the THAI project, our simulations were run with the UM's gravity wave drag scheme switched on, resulting in some differences in the wind structure.

All the simulations correspond to tidally locked planets. The Control ProxB, Warm ProxB, Control TRAP-1e, and Dry TRAP-1e simulations were run until a balance between incoming and outgoing radiation at top-of-atmosphere was achieved. Control ProxB, Warm ProxB, and Control TRAP-1e ran for 6,000 days and the period from day 5,000 to 6,000 was sampled for analysis. The Warm TRAP-1e simulation underwent a runaway greenhouse effect, with convection reaching the model top after approximately 4,000 days. We sampled a 990-day period (day 3,000 to the crash just before day 4,000) and include the results here to study the extreme limit of the habitable zone and in particular the potential effect of cloud variability on observations of close-in rocky planets (Venus analogues). As the Dry TRAP-1e simulation achieved radiative balance faster than the moist atmospheres, we ran the simulation for 4,000 days and used the period from day 3,000 to 4,000 for analysis. In the results reported below,``day 10'' and similar formulations refer to the day of the sample period, not the day of the simulation.

We performed one sensitivity test to investigate the effect of slab ocean depth on the period and amplitude of the variability. The slab ocean could potentially affect atmospheric variability because sea surface temperature is a direct forcing of the atmospheric circulation. We repeated the Control TRAP-1e simulation with a 50 m slab ocean instead of 1.0 m. We found no significant differences in the period and amplitude of the cloud cover oscillation described below and negligible differences in the climatology.

\subsection{NASA Planetary Spectrum Generator} \label{subsec:psg}
We use the NASA Planetary Spectrum Generator \citep{villanueva_planetary_2018}, publicly available at \url{https://psg.gsfc.nasa.gov/} (PSG), to simulate time series of water vapor and carbon dioxide features of the four moist atmosphere simulations as observed by the JWST's NIRSpec (Near Infrared Spectrograph) instrument. We omit the dry case as it has no time-varying atmospheric chemistry or clouds. The NIRSpec instrument's range covers water vapor features in the infrared at 1.4, 1.8, and 2.7 $\mathrm{\mu m}$, as well as CO$_2$ features at 2.1, 2.7, and 4.3 $\mathrm{\mu m}$ \citep{ahrer_identification_2023}. For each simulated atmosphere, we prescribe the orbital, planetary, and stellar parameters shown in Table \ref{tab:parameters}, together with the pressure, temperature, altitude, H$_2$O, N$_2$, and CO$_2$ data from the UM. For the Proxima Centauri b simulations, the 85 km model top was sufficient for the PSG to calculate a spectrum, and we use only the model output data. For the TRAPPIST-1e simulations, however, the 80 km model top was slightly too low to enable the PSG to model the spectrum. We used the Met Office's \emph{iris} package's built-in linear extrapolation method to extend the temperature, H$_2$O, ice cloud, liquid cloud, N$_2$ and CO$_2$ profiles to one extra atmospheric level with an altitude of 85 km and half the pressure of the layer immediately below \citep{metoffice_iris_2010}. Previous works have used the PSG to generate a spectrum for each grid box and averaged the spectra for a final output representing the signal during transit \citep{may_water_2021, suissa_dim_2020, komacek_clouds_2020}. To reduce the computational expense of simulating long time series for multiple simulations and absorption features, we instead average the atmospheric values for each day around the limb first, generate a transit spectrum for each day, and extract and plot the absorption features against time.

\section{Results} \label{sec:results}

\subsection{Climatology} \label{subsec:climate}
\begin{figure}[ht!]
\epsscale{1.0}
\plotone{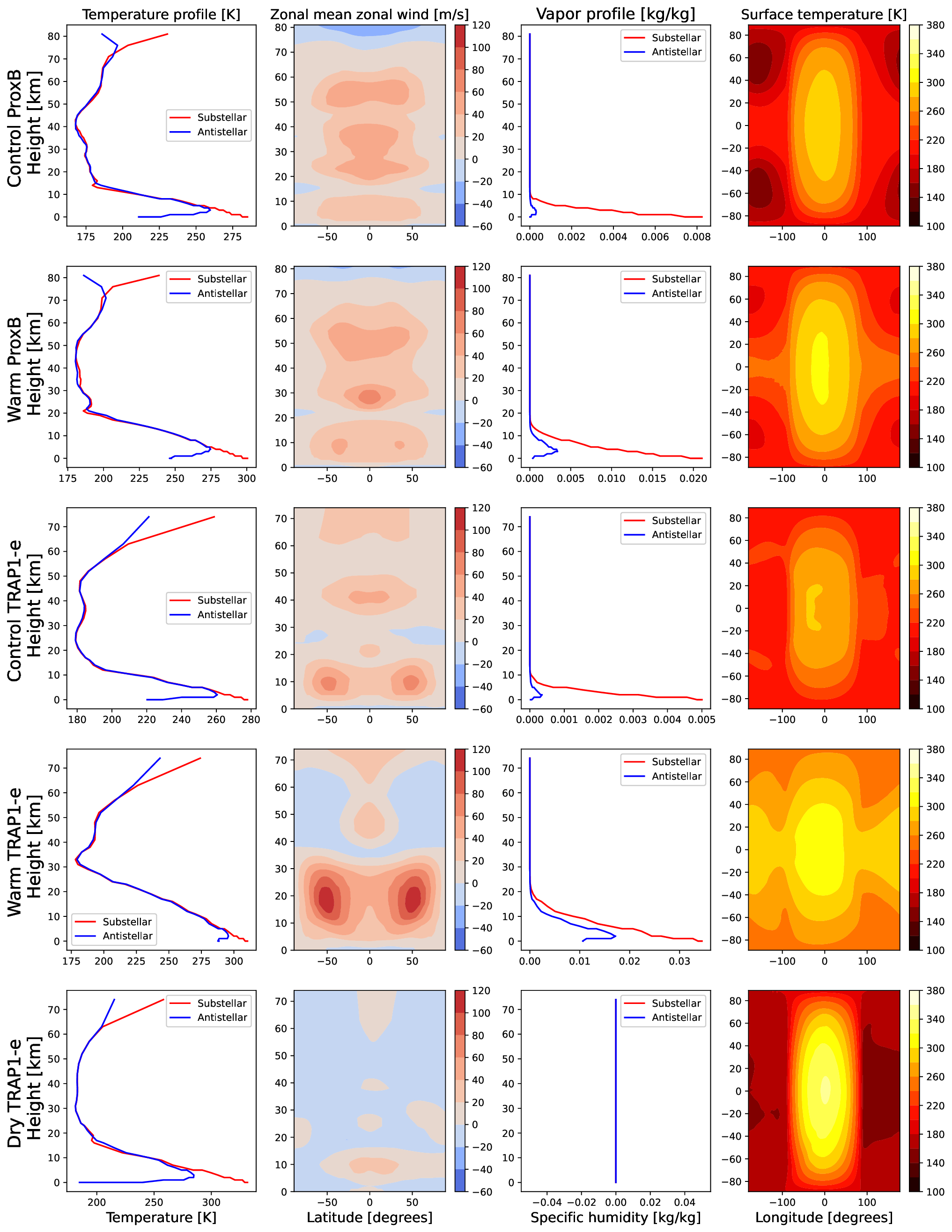}
\caption{Comparative climatology of the five simulations, showing (from left to right) the vertical temperature profile at the substellar and antistellar point, the zonal mean zonal wind, the vertical water vapor profile at the substellar and antistellar point, and the surface temperature. From top row to bottom row: Control ProxB, Warm ProxB, Control TRAP-1e, Warm TRAP-1e, Dry TRAP-1e. All values are 300-day means. 
\label{fig:climatology}}
\end{figure}

\begin{table}
\hspace*{-2cm} 
\begin{tabular}{llllll}
\hline
Quantity & Control ProxB & Warm ProxB & Control TRAP-1e & Warm TRAP-1e & Dry TRAP-1e  \\
\hline \hline
Mean air temperature (sub) (K) & 211.4 & 229.9  & 217.9 & 250.7 & 240.3 \\
Max air temperature (sub) (K) & 285.1 & 300.6 & 278.0 & 310.5 & 331.6 \\
Min air temperature (sub) (K) & 167.8 & 180.6 & 179.2 & 178.2 & 181.4 \\
Mean air temperature (anti) (K) & 201.6 &  220.6 & 210.4 & 246.3 & 222.1 \\
Max air temperature (anti) (K) & 259.7 & 274.7 & 260.6 & 295.8 & 284.7 \\
Min air temperature (anti) (K) & 167.9 & 180.4 & 179.0 &  179.4 & 181.4 \\
Mean zonal mean zonal wind (m/s) & 17.1 & 18.4 & 15. 6 & 25.2 & -1.2 \\
Max zonal mean zonal wind (m/s) & 59.0 & 73.5 & 69.6 & 113.9 & 24.8\\
Min zonal mean zonal wind (m/s) & -35.4 & -24.3 & -6.4 &  -17.4 & -20.1 \\
Mean specific humidity (sub) (kg/kg) & 1.5$\times$10$^{-3}$ &  4.9$\times$10$^{-3}$ &  1.0$\times$10$^{-3}$ &  10.5$\times$10$^{-3}$ & 0 \\
Max specific humidity (sub) (kg/kg) & 8.2$\times$10$^{-3}$ & 21.0$\times$10$^{-3}$ & 5.0$\times$10$^{-3}$ & 34.5$\times$10$^{-3}$ & 0 \\
Min specific humidity (sub) (kg/kg) & 0.5$\times$10$^{-7}$ & 4.8$\times$10$^{-7}$ & 0.1$\times$10$^{-7}$ & 2.5$\times$10$^{-7}$ & 0 \\
Mean specific humidity (anti) (kg/kg) & 0.4$\times$10$^{-4}$ & 6.2$\times$10$^{-4}$ & 0.6$\times$10$^{-4}$ & 52.9$\times$10$^{-4}$ & 0 \\
Max specific humidity (anti) (kg/kg) & 3.0$\times$10$^{-4}$ & 33.9$\times$10$^{-4}$ & 3.5$\times$10$^{-4}$ & 171.8$\times$10$^{-4}$ & 0\\
Min specific humidity (anti) (kg/kg) & 0.5$\times$10$^{-7}$ & 5.3$\times$10$^{-7}$ & 0.1$\times$10$^{-7}$ & 2.4$\times$10$^{-7}$ & 0 \\
Mean surface temperature (K) & 219.1 & 243.4 & 231.0 & 276.2 & 212.9 \\
Max surface temperature (K) & 286.3 & 301.8 & 281.6 & 312.3 & 342.5  \\
Min surface temperature (K) & 149.5 & 196.5 & 205.2 & 250.4 & 157.3 \\
\hline
\end{tabular}
\caption{Mean, maximum, and minimum values for each plot shown in Figure \ref{fig:climatology}. Values are given separately for the substellar and antistellar profiles of temperature and specific humidity.}
\label{tab:climstats}
\end{table}

\cite{boutle_exploring_2017} and \cite{sergeev_atmospheric_2020} present a detailed climatology of Proxima Centauri b as simulated by the UM. Similarly, a full description of the climatology of TRAPPIST-1e as simulated by the UM with a dry and moist atmosphere is given in \cite{turbet_trappist-1_2022} and \cite{sergeev_trappist-1_2022}, respectively. We give a brief overview of the equilibrium climates of all five simulations here.

Figure \ref{fig:climatology} shows the vertical temperature structure, zonal mean zonal wind, vertical humidity profile, and the spatial distribution of surface temperature of each simulation. All simulations display a nightside temperature inversion, although in the Warm TRAP-1e case it is very small and the temperature profile is nearly identical on the dayside and nightside. The specific humidity profiles are likewise consistent for Control ProxB, Warm ProxB, and Control TRAP-1e, with much greater humidity on the dayside and an arid nightside. Only the Warm TRAP-1e (incipient runaway) simulation has substantial humidity on the nightside. A comparison of the zonal mean zonal wind of the Control vs. Warm ProxB and Control vs. Warm TRAP-1e cases supports an increase in zonal wind speeds for planets orbiting closer in. The Proxima Centauri b simulations have a broad equatorial jet in the troposphere and a series of vertically stacked opposing jets in the stratosphere in a longitudinally asymmetric stratospheric oscillation (LASO) as described in \cite{cohen_longitudinally_2022}. In contrast, the Control and Warm TRAPPIST-1e simulations form a mid-latitude tropospheric jet in each hemisphere. In these simulations, unlike in the THAI project, the planet also generates a LASO in the equatorial region due to the acceleration of the flow contributed by the gravity wave drag scheme. 

The zonal mean zonal wind for the Dry TRAP-1e case differs from that reported in THAI \citep{turbet_trappist-1_2022} for the equivalent N$_2$-dominated atmosphere case. \cite{turbet_trappist-1_2022} reported a stable state with two mid-latitude jets, while our result is a broad equatorial jet more similar to that of Proxima Centauri b or the CO$_2$-dominated atmosphere case in \cite{turbet_trappist-1_2022}. Recent work has shown that UM simulations of TRAPPIST1-e exhibit climate bistability, with one stable dynamical state corresponding to an equatorial jet and the other to two mid-latitude jets \citep{sergeev_bistability_2022}. It may be that the inclusion of gravity waves, which affect the dynamical structure of the atmosphere and heat transport between dayside and nightside, tipped this simulation into the equatorial jet state. Using a series of daily snapshots, we found that our Dry TRAP-1e simulation had considerably greater day-to-day wind variability in the troposphere than the equivalent publicly available THAI Ben1 simulation that lacked gravity waves. This increase in variability is likely due to non-linear interaction between the gravity wave scheme and other elements of the circulation. The long-term mean horizontal flow was, however, very similar between the two simulations, including the positions of the Rossby gyres, with the primary difference being the lack of fast mid-latitude jets in the mean. The zonal wind magnitude for the Dry TRAP-1e case is in line with that shown in \cite{turbet_trappist-1_2022} and considerably less than that in the moist atmosphere cases. The tropospheric jet structure influences the location and shifting of Rossby wave structures discussed below, as the zonal wind magnitude is a component of the Rossby wave phase speed and Rossby waves can be advected by the flow.

\subsection{Wave oscillation and mechanism} \label{subsec:mechanism}
\begin{figure}[ht!]
\gridline{\fig{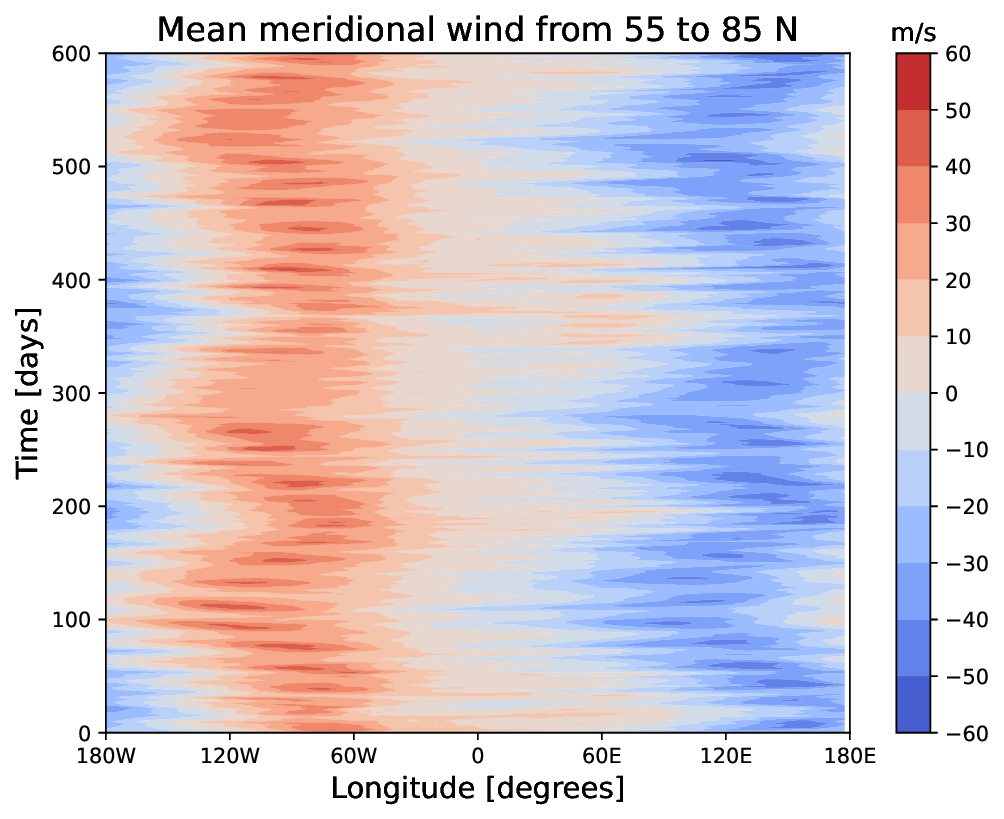}{0.45\textwidth}{a) Control ProxB}
          \fig{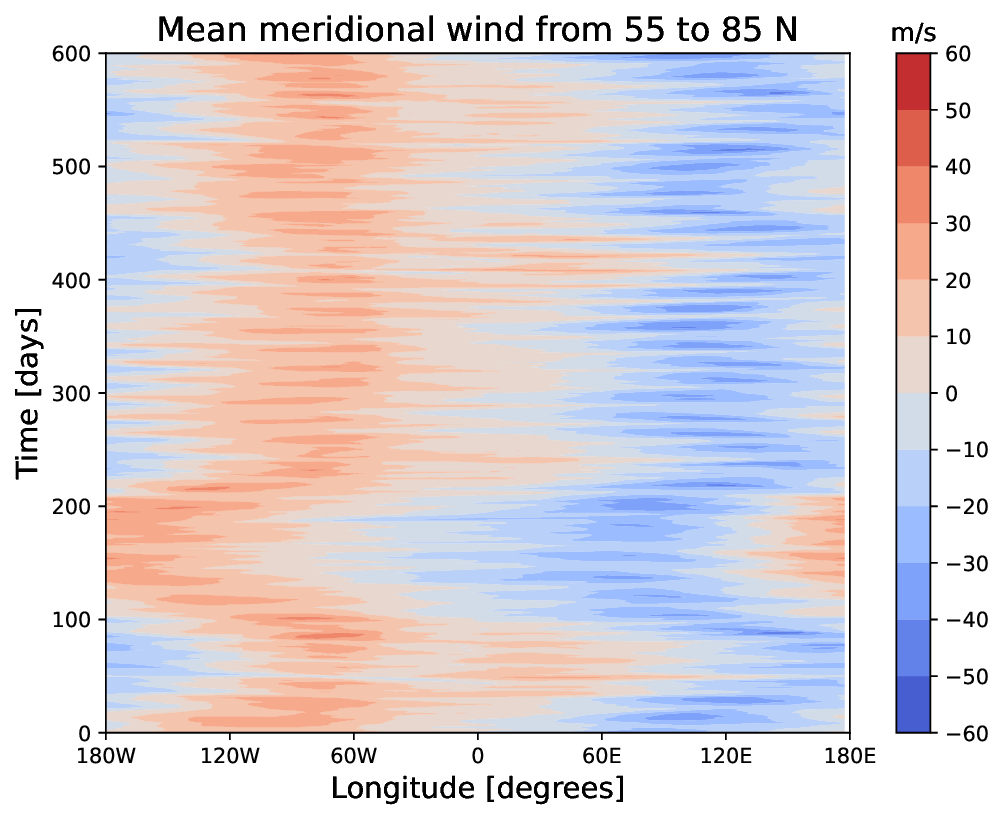}{0.45\textwidth}{b) Warm ProxB}}
\gridline{\fig{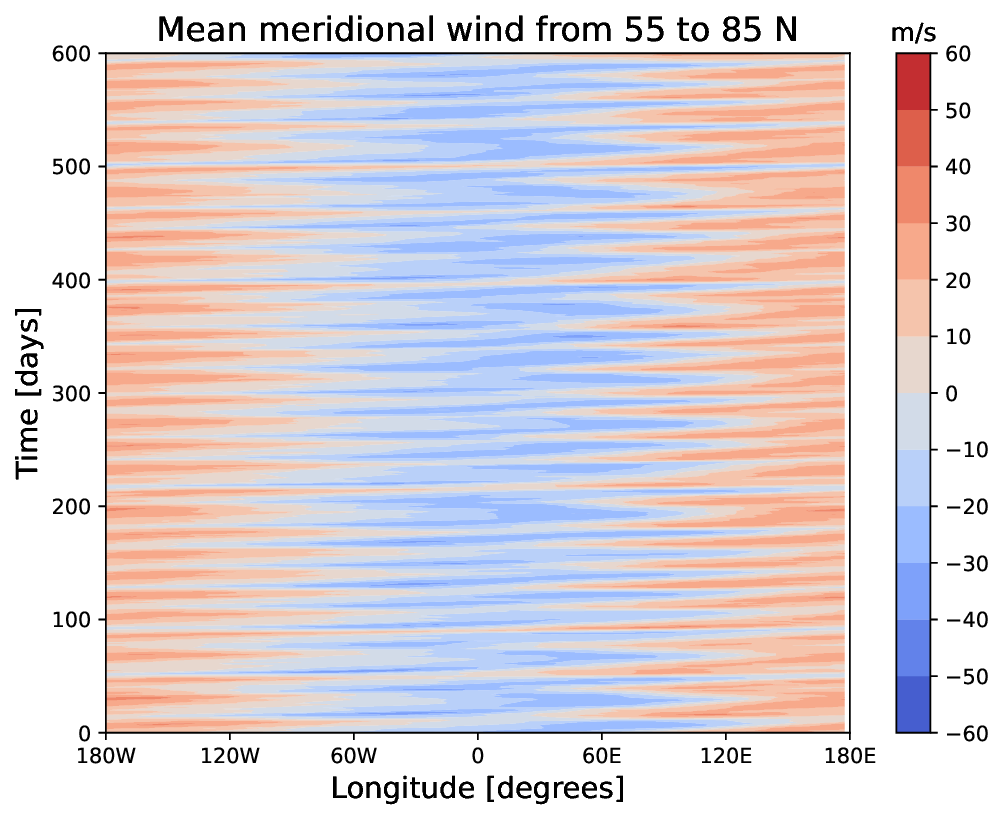}{0.45\textwidth}{c) Control TRAP-1e}
          \fig{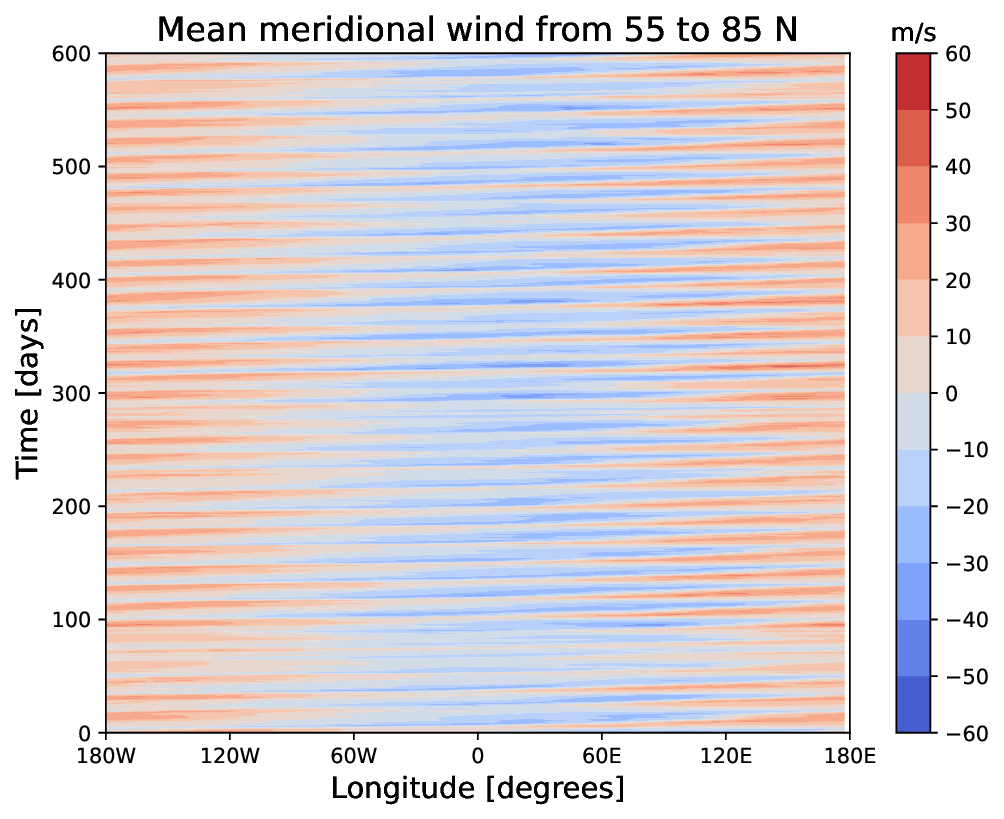}{0.45\textwidth}{d) Warm TRAP-1e}}
\gridline{\fig{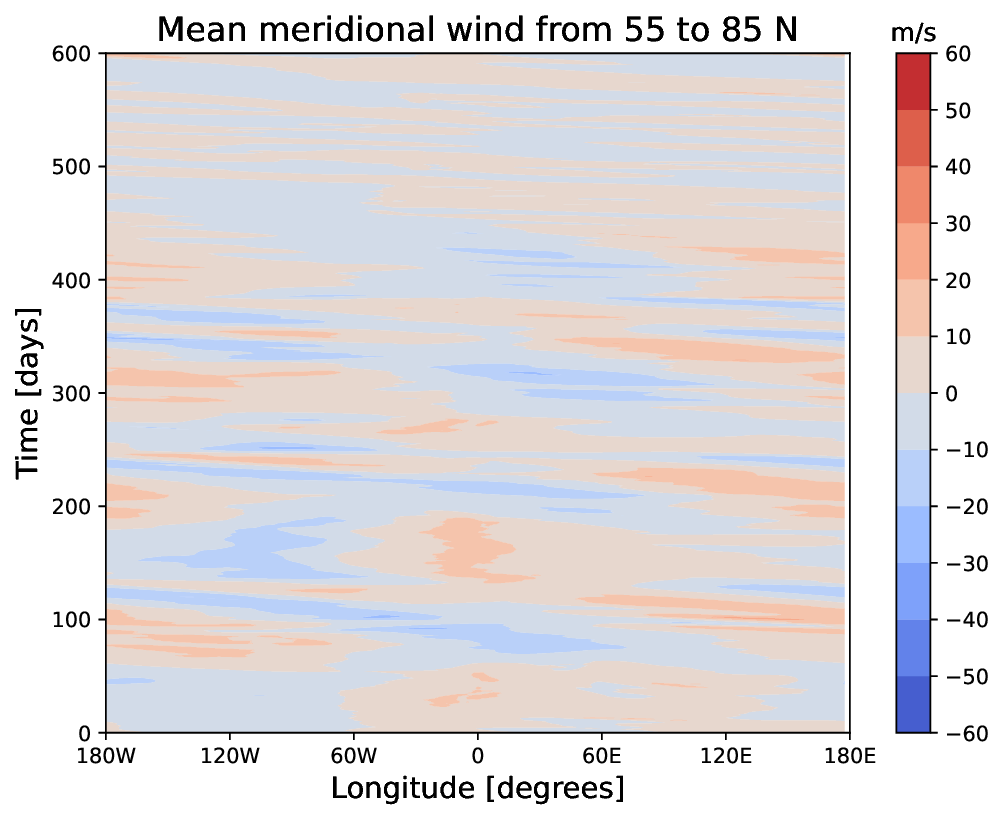}{0.45\textwidth}{e) Dry TRAP-1e, high latitudes}
          \fig{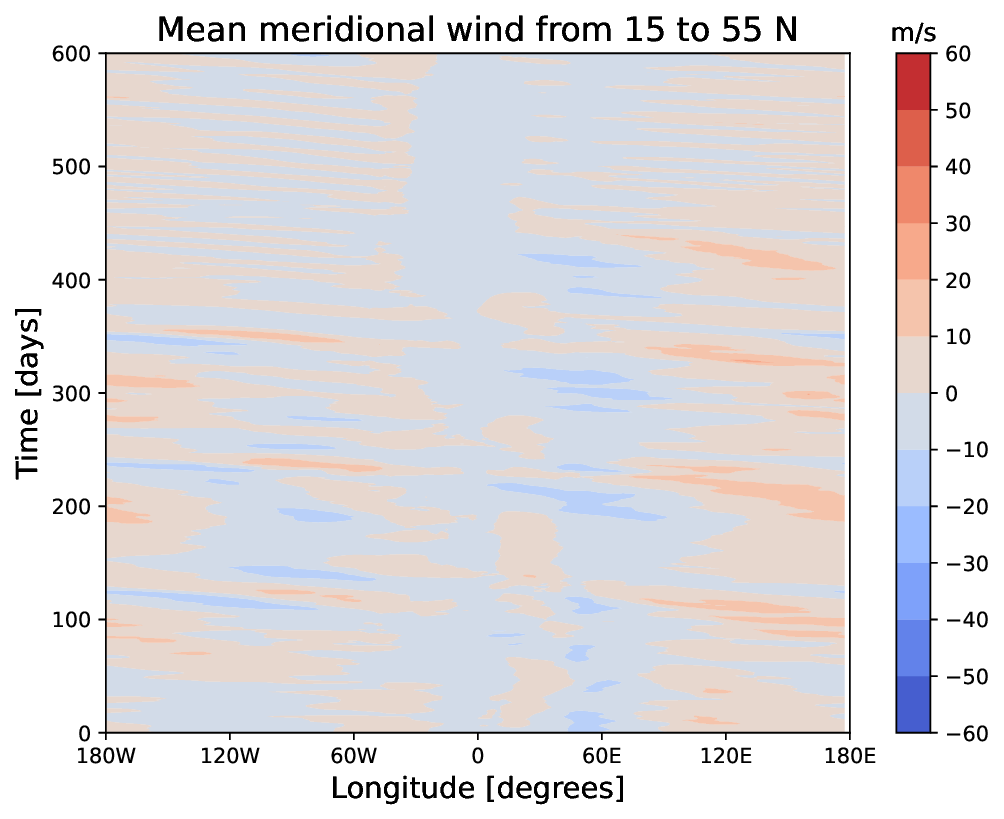}{0.45\textwidth}{f) Dry TRAP-1e, low latitudes}}
\caption{Time-longitude diagrams of the mid-latitude (55 to 85N) averaged meridional wind at an altitude of 2.96 km above the surface. Subplot f) also shows the low latitudes for Dry TRAP-1e. Positive values correspond to northward flow, while negative values represent southward flow.}
\label{fig:rwaves}
\end{figure}

All five simulations exhibit the presence of zonal wavenumber-1 Rossby waves. In all the moist cases, these waves shift eastwards and westwards around an equilibrium position with a regular period. The zonal wavenumber-1 Rossby wave response arises due to the spatially periodic heating caused by irradiation of the permanent dayside of a tidally locked planet and lack of irradiation of the nightside \citep{showman_equatorial_2011, showman_matsuno-gill_2010, gill_simple_1980, matsuno_quasi-geostrophic_1966}. Dayside heating causes a region of wind divergence around the substellar point in the upper atmosphere \citep{hammond_rotational_2021}. A region of divergence superimposed on an absolute vorticity gradient creates a Rossby wave source as described in \cite{sardeshmukh_generation_1988} in the substellar region. Past work has reported that the eastward flow in tidally locked planet simulations causes a phase shift in the zonal wavenumber-1 Rossby wave response dependent on the long-term mean zonal wind speed \citep{wang_phase_2021, hammond_wave-mean_2018}. We observe this long-term mean phase shift in our simulations, but find that in a time-resolved analysis, the wave response shift (i.e., location of the Rossby minima and maxima) varies periodically. This wave oscillation induces regular cloud cover drops at the eastern terminator in two of our simulations: Control ProxB and Warm ProxB.

The wave shift causes cloud cover variations in the Proxima Centauri b simulations and not the TRAPPIST-1e simulations because the Rossby waves form in the mid-latitudes and therefore interact with the substellar cloud. The gyres in Control ProxB and Warm ProxB are centered around 45-60N/S and extend to nearly the poles and equator, while the substellar cloud region reaches from 60S to 60N. In Control TRAP-1e, the eastern gyres are centered around 60-70N/S, while the clouds only extend to roughly 30N/S: the latitude range of the gyres does not overlap with that of the substellar cloud. In Warm TRAP-1e, the bulk of the cloud forms higher up in the atmosphere and equatorwards of 15N/S, while the gyres are again located at 60-70N/S. The latitudinal position of the gyres may be influenced by the position of the zonal jets on the two planets. As seen in Figure \ref{fig:climatology}, Proxima Centauri b's equatorial jet extends to about 55N/S, while the mid-latitude jets on TRAPPIST-1e extend further polewards to approximately the latitude of the Rossby vortices. Previous work simulating TRAPPIST-1e with the UM has also shown that the circulation of this planet can take on one of two regimes: a single equatorial jet or two mid-latitude jets \citep{sergeev_bistability_2022}. This study found that in the single equatorial jet regime, as in our Proxima Centauri b simulations, the gyres form further equatorwards than in the double mid-latitude jet regime.

\begin{figure}[ht!]
\gridline{\fig{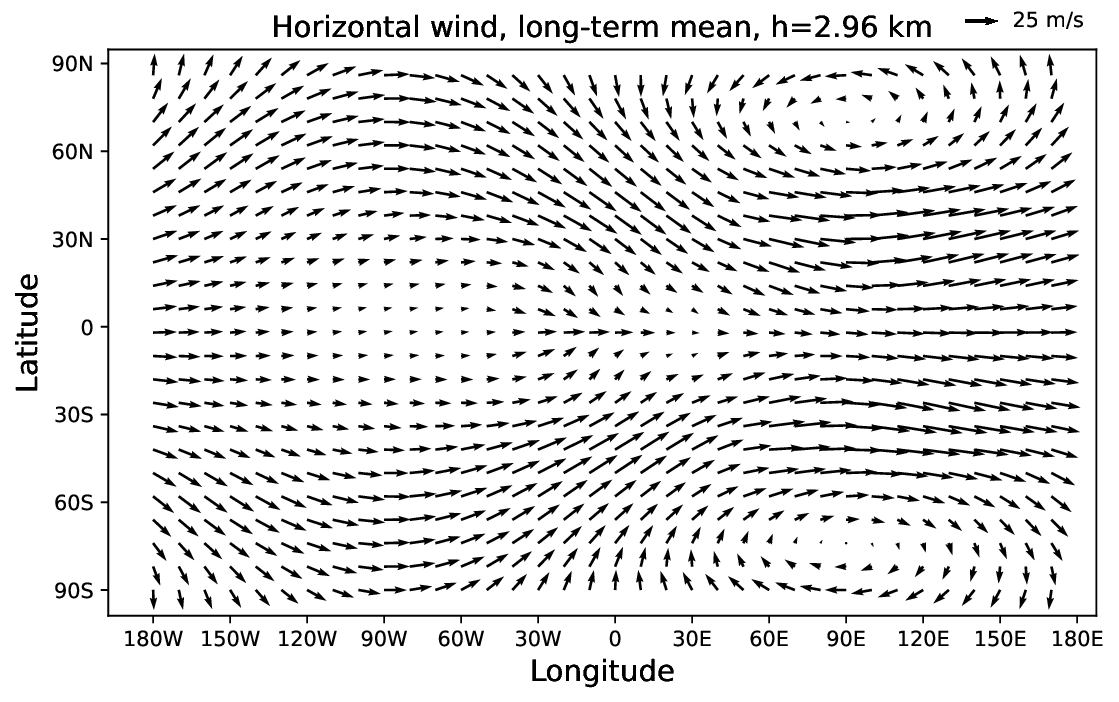}{0.45\textwidth}{a) Control TRAP-1e, long-term mean}
            \fig{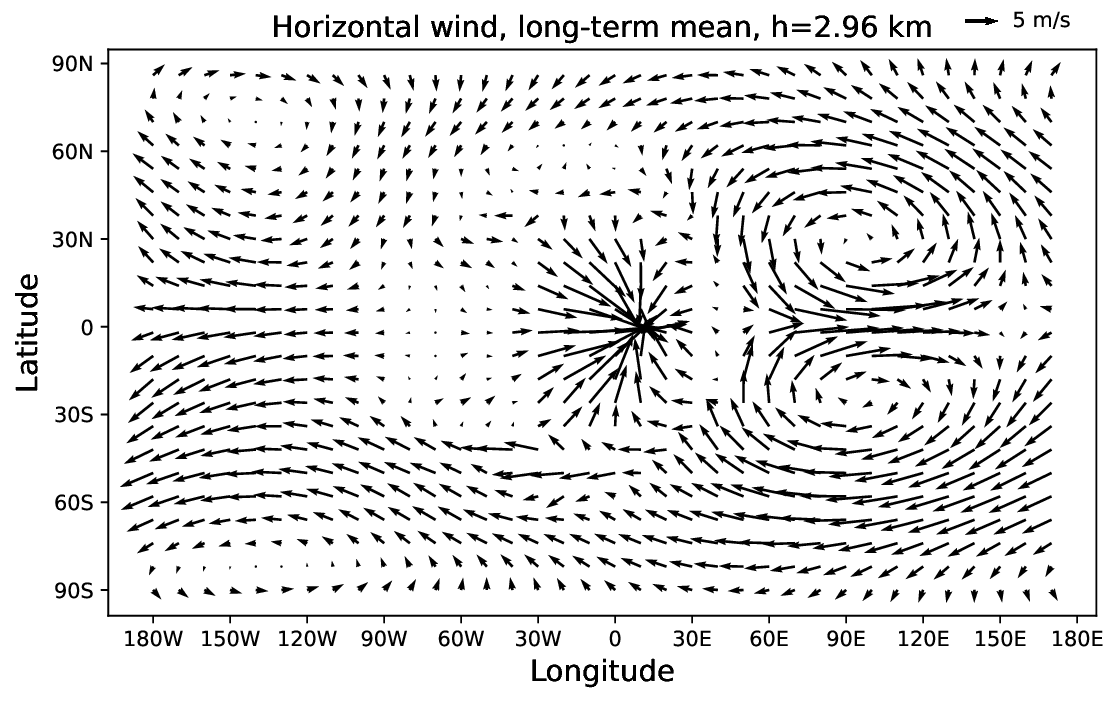} {0.45\textwidth}{e) Dry TRAP-1e, long-term mean}}
\gridline{\fig{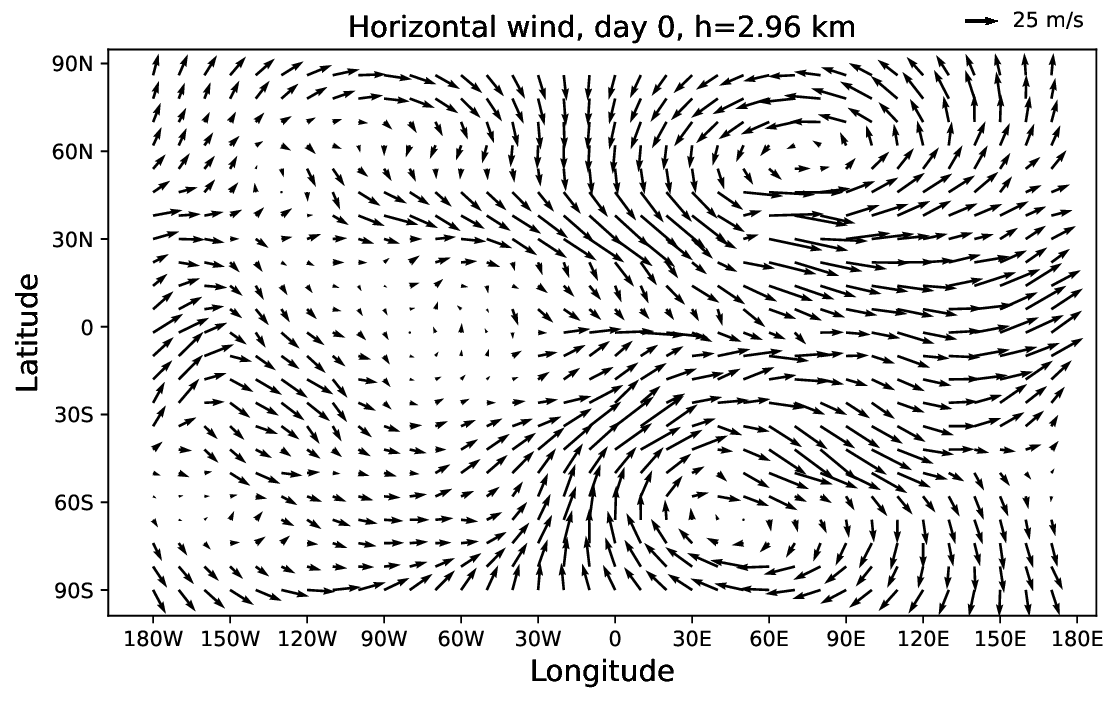}{0.45\textwidth}{b) Control TRAP-1e, day 0}
          \fig{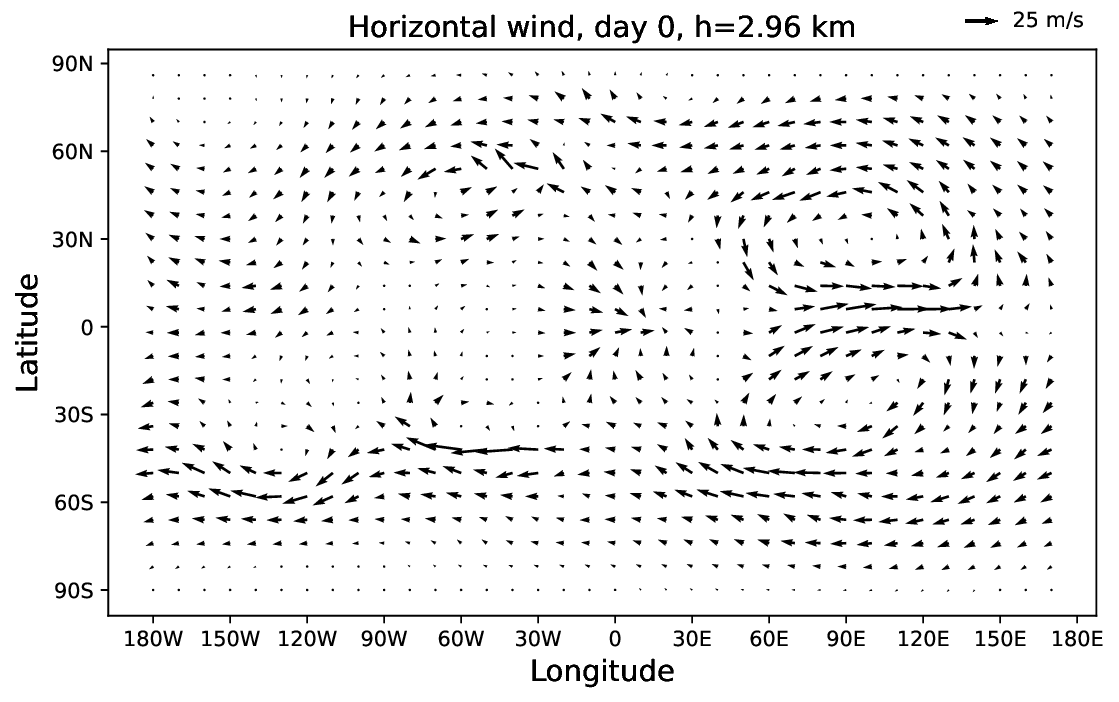}{0.45\textwidth}{f) Dry TRAP-1e, day 0}}
\gridline{\fig{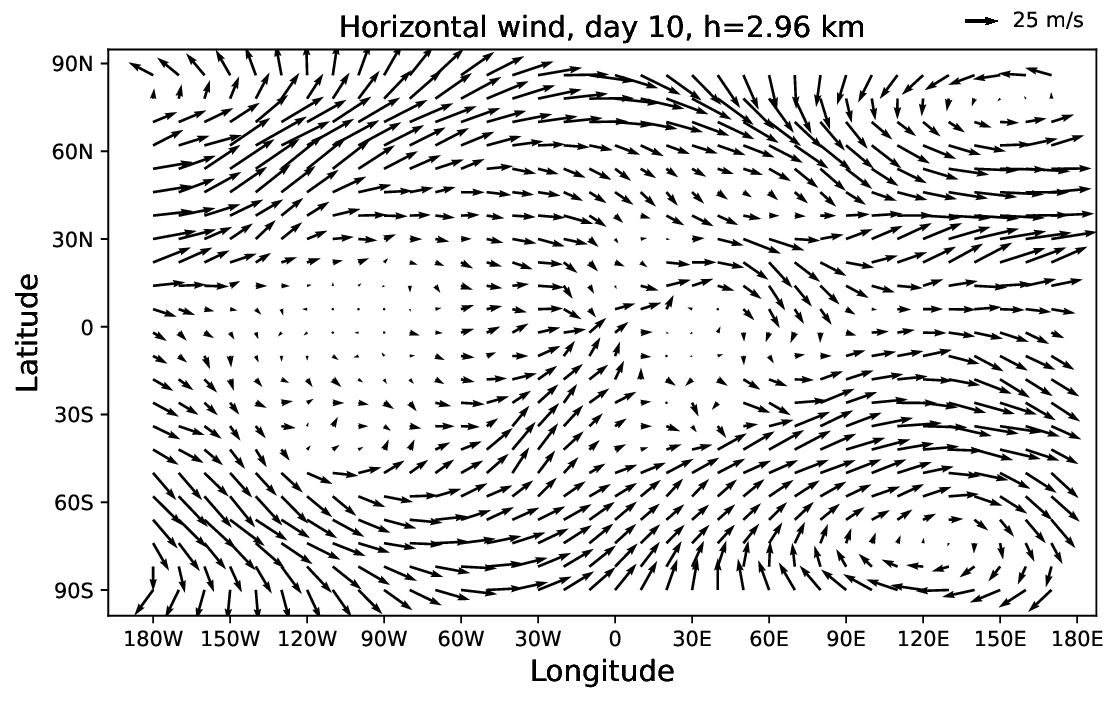}{0.45\textwidth}{c) Control TRAP-1e, day 10}
          \fig{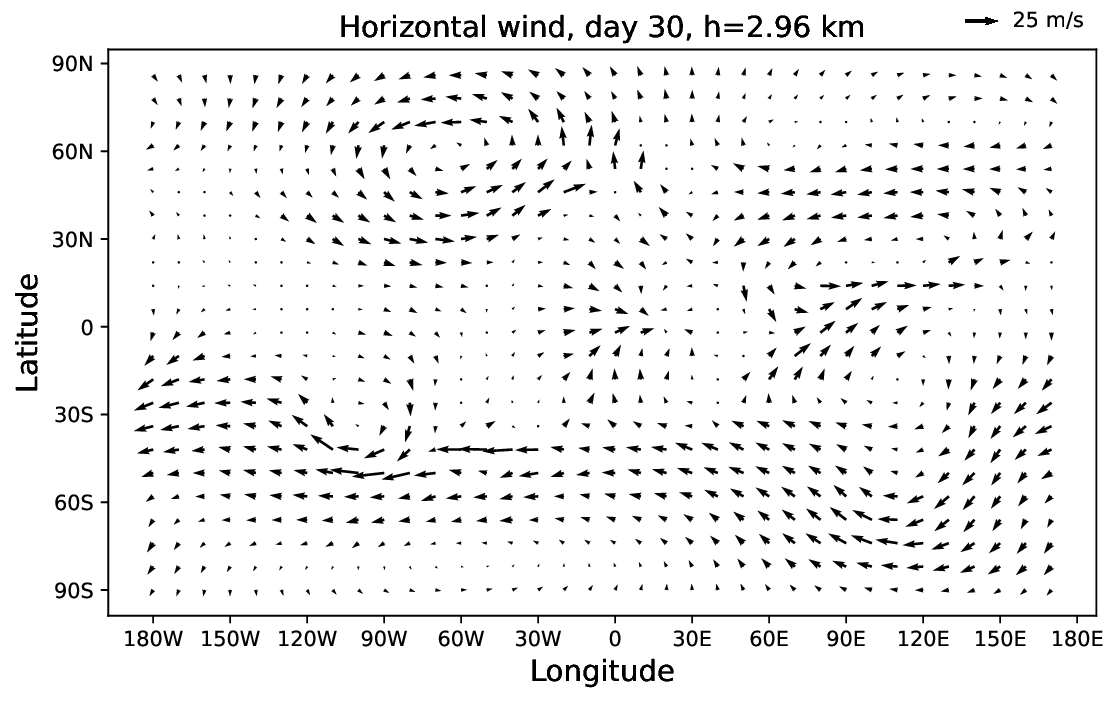}{0.45\textwidth}{g) Dry TRAP-1e, day 30}}
\gridline{\fig{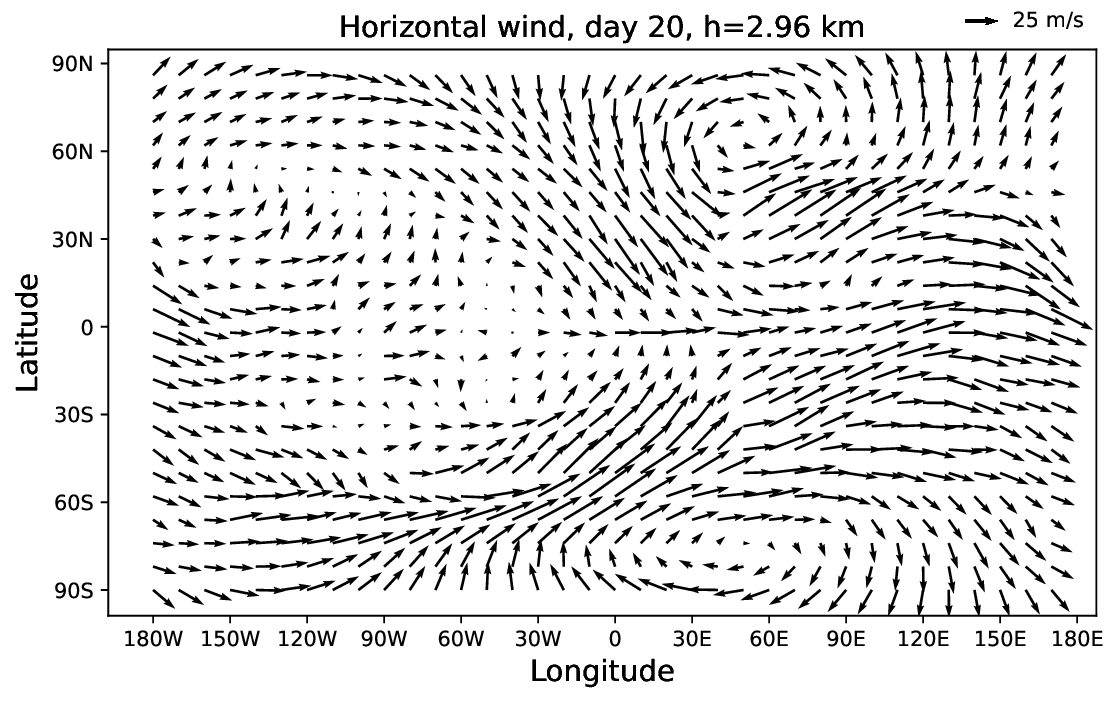}{0.45\textwidth}{d) Control TRAP-1e, day 20}
          \fig{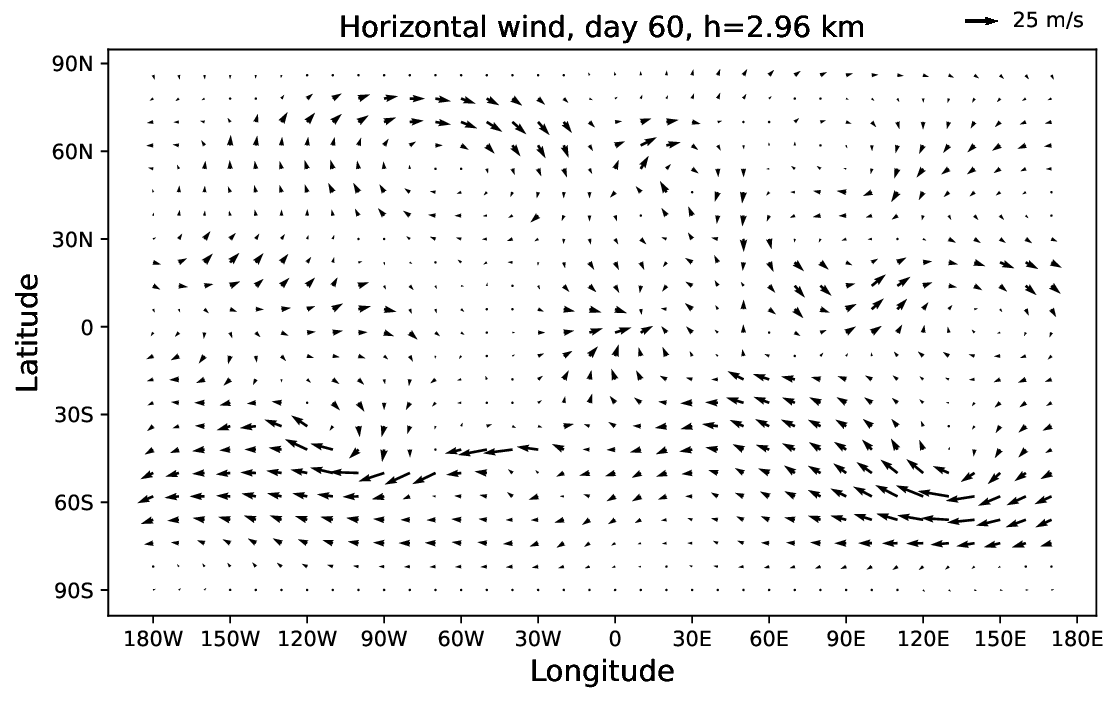}{0.45\textwidth}{h) Dry TRAP-1e, day 60}}
\caption{Long-term mean of the general circulation for Control TRAP-1e and Dry TRAP-1e, as well as daily snapshots of days 0, 10, and 20 and days 0, 30, and 60, respectively. Note the differing quiver scale in c).}
\label{fig:quivers}
\end{figure}

Figure \ref{fig:rwaves} represents the waves as fluctuations in the mean mid-latitude meridional wind at a height of 2.96~km. Investigation of the vertical vorticity profile and inspection of the eddy rotational component at different atmospheric levels showed that the Rossby wave-associated vorticity and wind speeds in the vertical region where clouds form are greatest at this height. Accordingly, single-level plots in our results are shown at 2.96~km. Results for other levels are qualitatively similar but typically of smaller magnitude. In Figure \ref{fig:rwaves} a) to d), the longitudes at which the mean meridional wind alternates between northward and southward represent the longitudinal range of the oscillation in the moist atmosphere cases. The Dry TRAP-1e case is shown in Figure \ref{fig:rwaves} e) and f). As is visible in the long-term mean flow of Dry TRAP-1e depicted in Figure \ref{fig:quivers} e), the western gyres in this simulation form at mid-to-high latitudes (60-90N/S), while the eastern gyres form at low latitudes (0-45N/S). The latitude range chosen in Figure \ref{fig:rwaves} e) includes the western gyres, which tend to propagate west, dissolve, and reform near the substellar longitude, but occasionally also propagate eastwards or remain stationary. The eastern gyres are always stationary: their longitudinal position can be seen at 90E in Figure \ref{fig:rwaves} f). To understand the mechanism driving the moist oscillation, we analyzed and compared the Control and Dry TRAP-1e simulations. 

Figure \ref{fig:quivers} shows the wind pattern at 2.96~km for Control TRAP-1e and Dry TRAP-1e in the long-term mean and at three different simulation times, chosen to correspond to the easternmost, westernmost, and again easternmost location of the gyres in the Control TRAP-1e simulation, covering a full cycle of motion. In the control simulation, Rossby gyres are clearly visible in the northern and southern polar regions of the eastern hemisphere, for example at 60N and 60-90E in Figure \ref{fig:quivers} b). These gyres propagate eastwards and westwards such that the centers of the gyres shift from between 30-60E to 120-150E on an approximately 20-day cycle, with a long-term mean position of 85E. A matching western pair is less apparent due to interactions with other elements of the flow. In the dry simulation, the eastern gyres form at lower latitudes and remain stationary, while the western gyres propagate exclusively westwards, dissolve, and reform near the substellar longitude. 

We explain the motion of the gyres using the theory of Rossby waves. Following e.g., \cite{holton_introduction_2013} or \cite{vallis_atmospheric_2017}, the phase speed of travelling Rossby waves is given by:

\begin{equation}
    c_p = \bar U - \frac{\beta + \bar U k_d^2}{k^2 + l^2 + k_d^2},
\label{eqn:cp}
\end{equation}

where \emph{$\bar U$} is the zonal mean zonal wind speed, $\beta$ is the Rossby parameter $\frac{2 \Omega cos \phi}{r}$ (with $\Omega$ the planet's rotation rate in radians/second, $\phi$ the latitude in radians, and r the planet's radius), and \emph{k} and \emph{l} are the zonal and meridional wavenumbers in units of m$^{-1}$. The variable $k_d$ is the wavenumber corresponding to the Rossby deformation radius $\frac{1}{L_d}$ rather than the planet's radius,  where $L_d$ is defined in quasigeostrophic theory as $\frac{NH}{f_0}$, with N the Brunt-V{\"a}is{\"a}l{\"a} frequency, H the scale height (6800 m for our simulations), and $f_0$ the Coriolis parameter at a given latitude. This form of the Rossby wave phase velocity equation takes into account vertical stratification and propagation within a GCM. For stationary Rossby waves such as those in our simulations, the theoretical phase speed corresponds to a longitudinal shift in the wave response location as the waves are advected by the zonal flow.

\begin{figure}[ht!]
\plotone{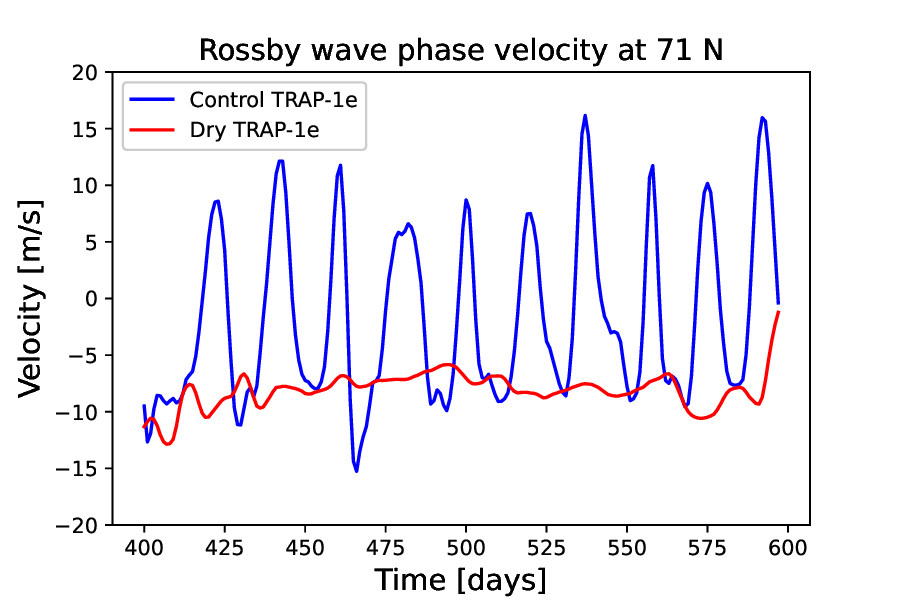}
\plotone{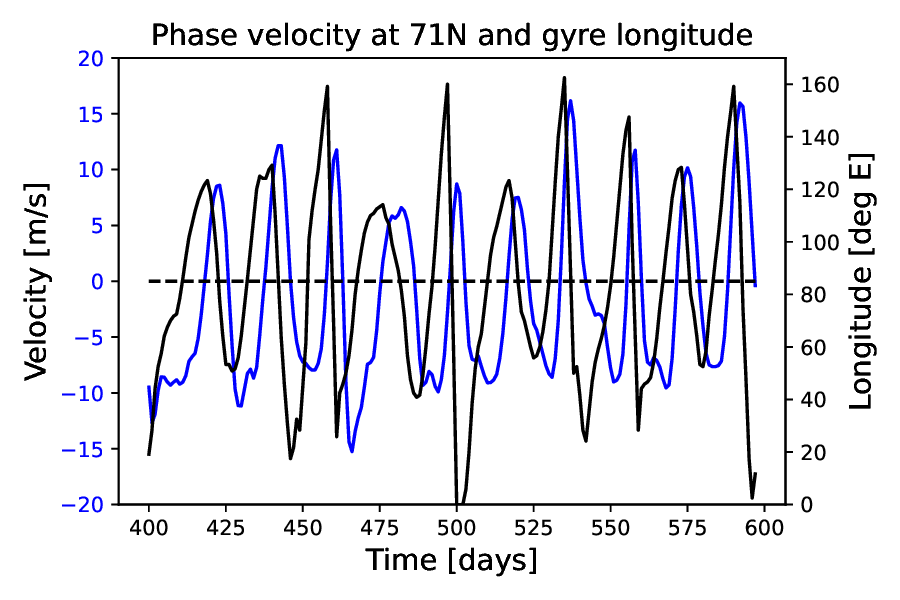}
\caption{Top: Time series of Rossby wave phase velocity at 71N for Control and Dry TRAP-1e simulations. Bottom: Time series of Rossby wave phase velocity at 71N overlaid with the longitudinal position of the northeast Rossby gyre for Control TRAP-1e. A 3-day rolling mean has been applied to all curves.}
\label{fig:rossbycp}
\end{figure}

To determine why the Rossby gyres oscillate in the moist atmosphere cases only, we compared the Rossby wave phase velocity for Control TRAP-1e and Dry TRAP-1e. Figure \ref{fig:rossbycp} a) shows time series of the phase velocity of the Rossby wave with the highest power spectral density (PSD) in the flow, the zonal wavenumber-1 wave, for these two simulations. To confirm that only this wave contributes significantly to the phase speed, we extracted the wavenumbers of the highest powered waves in the flow on each simulation day. We first performed a Helmholtz decomposition of the wind field at 2.96~km to calculate the eddy rotational component as in \cite{hammond_rotational_2021}. We then input the magnitude of the eddy rotational component, which we equate to the Rossby waves, into a 2-D Fourier transform and extracted the zonal (k) and meridional (l) wavenumbers of the wave with the maximum PSD on each simulation day. Our results confirmed that the wave with k=1 and l=0 is consistently the highest powered wave. Finally, we calculated a day- and latitude-specific Rossby wave phase velocity as per Equation \ref{eqn:cp}. In this calculation, we used the time-varying daily value of the zonal mean zonal wind U and Brunt-V{\"a}is{\"a}l{\"a} frequency N, and further subtracted the long-term mean zonal mean wind to account for the long-term phase shift of the wave response described in \cite{wang_phase_2021} and \cite{hammond_wave-mean_2018}. Figure \ref{fig:rossbycp} a) shows that, in the moist simulation, the Rossby wave phase velocity oscillates between positive (eastward) and negative (westward) values on an approximately 20-day cycle, while it remains negative in the dry case over the same time period.

In Figure \ref{fig:rossbycp} b), we then plot the Rossby wave phase velocity for Control TRAP-1e as in a), together with the longitude of the center of the northeast Rossby gyre. We tracked the gyre center by searching for the longitude in the northeast quarter of the globe where the meridional wind changes direction for each simulation day. The dashed black line represents the equilibrium position of the gyre at 85E and is aligned with the zero point of the phase velocity on the plot. This plot shows the close and regular correlation between the gyre location at the given latitude and the Rossby wave phase speed at that same latitude in both period and amplitude. According to the interpretation that the phase speed of a stationary Rossby wave represents the wave's longitudinal position, a positive phase speed should correspond to a gyre longitude east of the equilibrium point. Our phase velocity curve conforms to this prediction except for a small, consistent offset from the gyre longitude curve. This offset is caused by the limitation of using the zonal wind at only one latitude in Eq. \ref{eqn:cp}, in particular a latitude at which the Rossby gyre itself also contributes to the zonal wind. Using the global mean zonal wind instead of the latitude-specific zonal wind in the phase velocity calculation reduces the offset between the two curves in Fig. \ref{fig:rossbycp} b) to nearly zero, but in turn weakens the correlation between the amplitudes of the curves. As the gyre extends over roughly 20 degrees latitude, treating it as a point particle with a single location is inadequate to precisely predict its motion; however, the strong correlation in period and amplitude in Fig. \ref{fig:rossbycp} b) supports the interpretation of the phase velocity as representative of the longitudinal shift in the wave response over time.

\begin{figure}[ht!]
\gridline{\fig{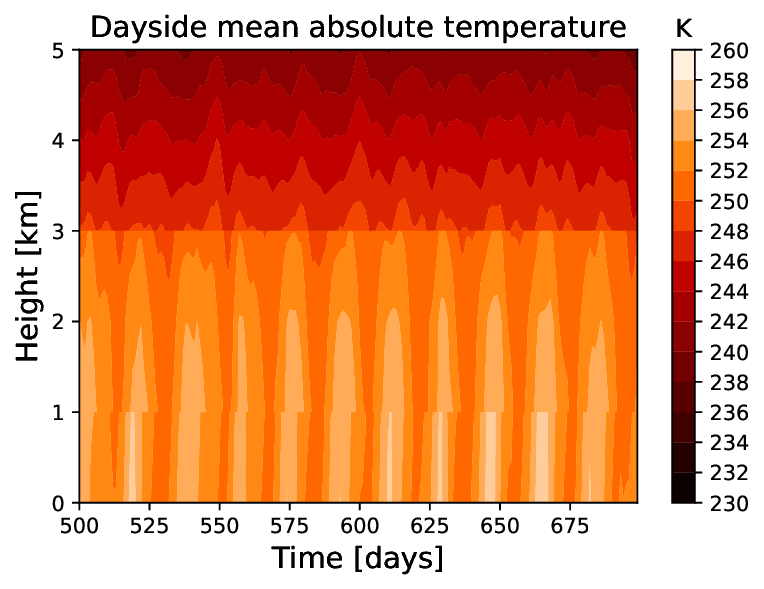}{0.35\textwidth}{a) Control TRAP-1e}
         \fig{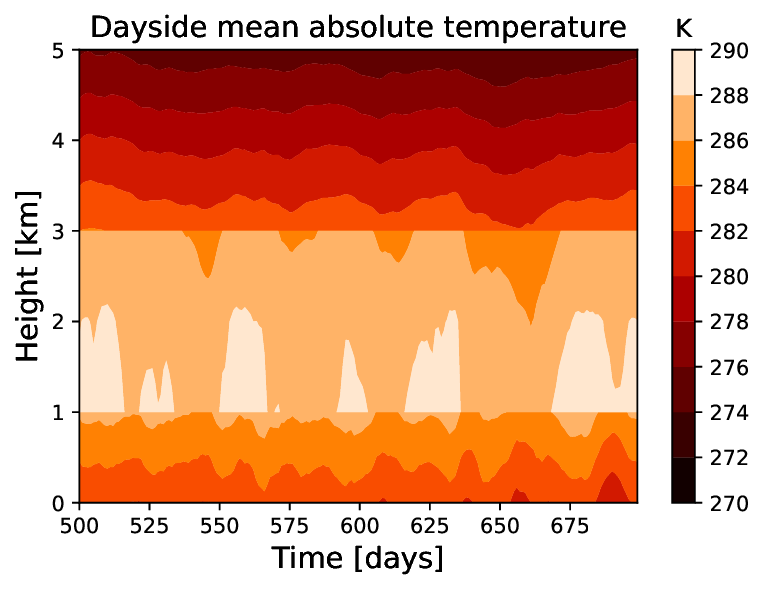}{0.35\textwidth}{b) Dry TRAP-1e}}
\gridline{\fig{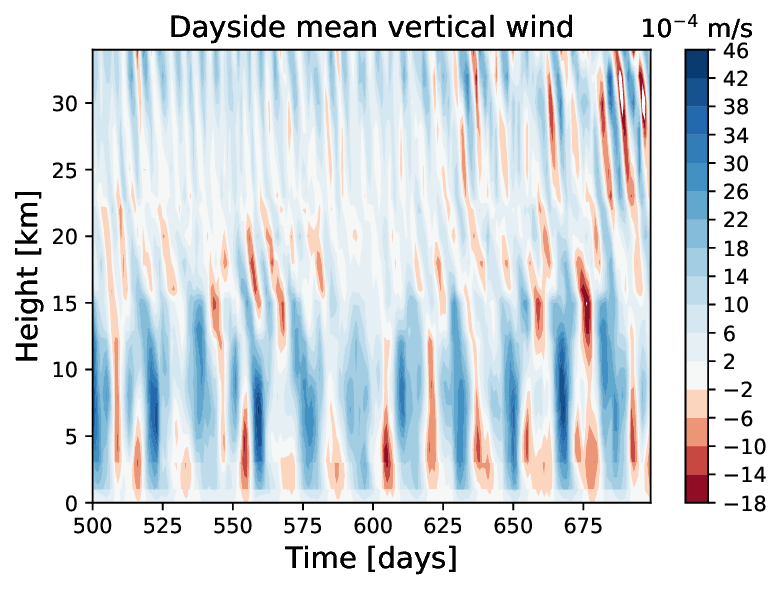}{0.35\textwidth}{c) Control TRAP-1e}
         \fig{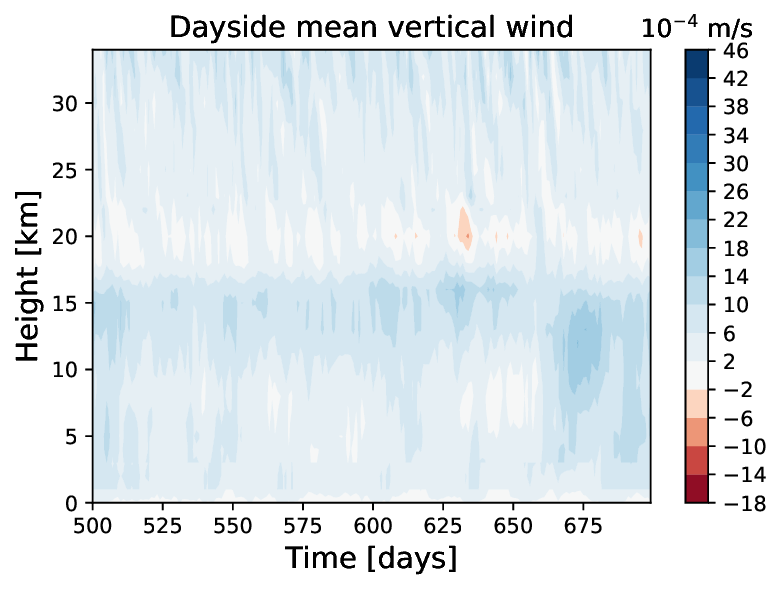}{0.35\textwidth}{d) Dry TRAP-1e}}
\gridline{\fig{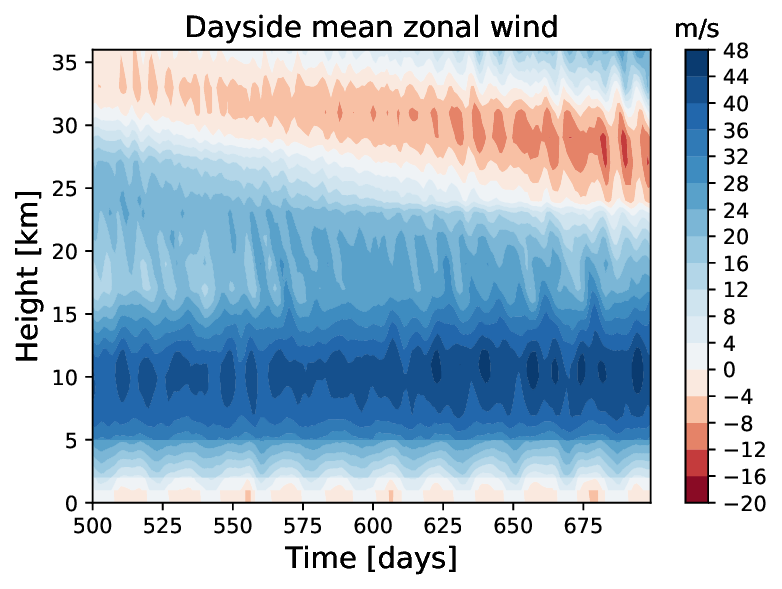}{0.35\textwidth}{e) Control TRAP-1e}
         \fig{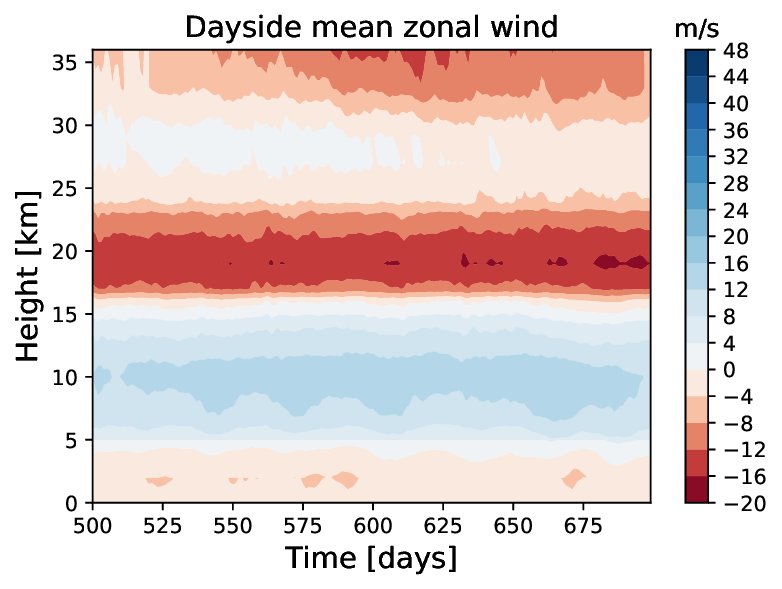}{0.35\textwidth}{f) Dry TRAP-1e}}
\gridline{\fig{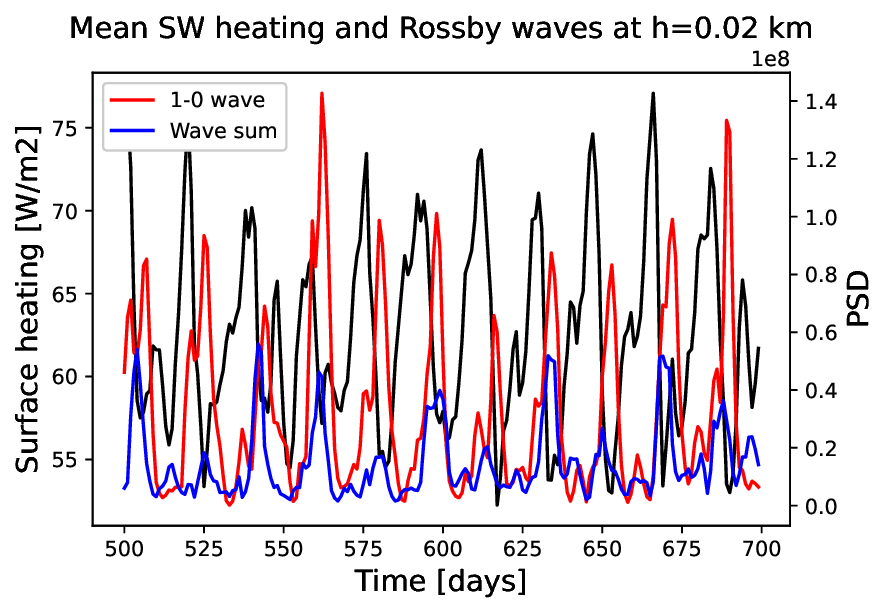}{0.35\textwidth}{g) Control TRAP-1e}
         \fig{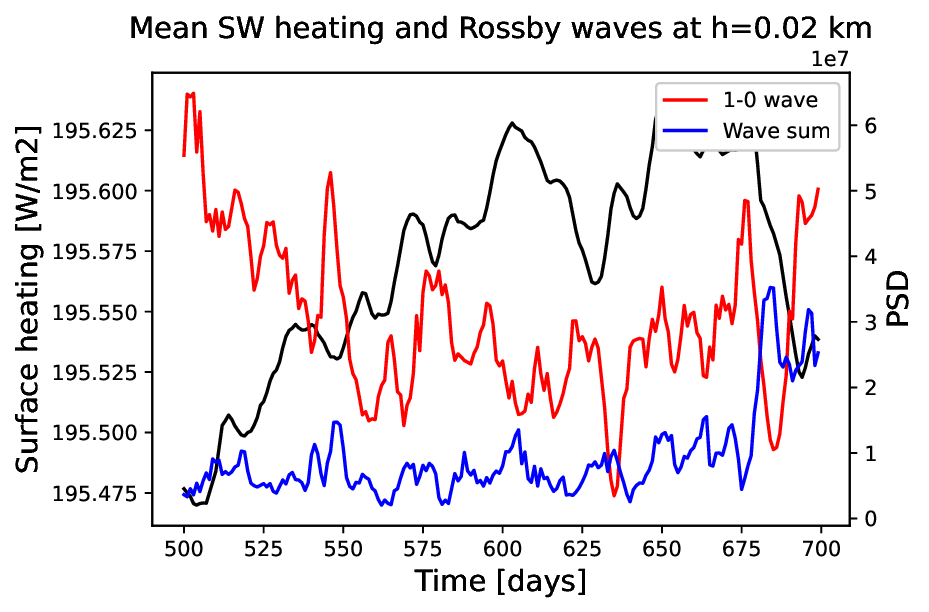}{0.35\textwidth}{h) Dry TRAP-1e}}
\caption{Top three rows: Vertical profiles of dayside mean air temperature, vertical wind, and zonal wind over time for the moist atmosphere Control TRAP-1e and the dry atmosphere Dry TRAP-1e. The vertical range of a) and b) is 0 to 5~km to better show the temperature oscillation near the surface. Due to the relatively low resolution of our simulations, this close-in view results in discontinuities between vertical levels. The discontinuities are not visible in c)-f) because the vertical range shown in 0 to 35~km. Bottom row: Time series of the dayside mean net downward shortwave flux close to the planet's surface (black), shown with the power spectral density of the zonal wavenumber 1 Rossby wave (red) and the sum of the power spectral density of the waves with zonal and meridional wavenumbers 1-1, 2-1, 2-2, and 3-2 (blue). Note the different limits of the y-axis.}
\label{fig:resonance}
\end{figure}

The formation of Rossby gyres in simulations of tidally locked planets is believed to be related to the spatially periodic thermal forcing \citep{showman_equatorial_2011, showman_matsuno-gill_2010, gill_simple_1980, matsuno_quasi-geostrophic_1966}. As our model's stellar spectra do not vary with time and the planet is tidally locked with zero eccentricity or obliquity, atmospheric processes must be responsible for the temporal variability in our simulations. To determine why the Rossby wave phase velocity varies periodically with time, we searched for correlations between quantities thought to play a role in the Matsuno-Gill response to periodic forcing. Figure \ref{fig:resonance} compares the variations over time of the vertical cross sections of dayside mean air temperature, vertical wind, zonal wind, net surface shortwave flux, and the PSD of the zonal wavenumber 1 Rossby wave (identified as the 1-0 wave) and, separately, the sum of the PSDs of the Rossby waves identified as 1-1, 2-1, 2-2, and 3-2 waves, where the first digit refers to the zonal wavenumber and the second digit refers to the meridional wavenumber. We show the 1-0 wave separately from other long Rossby waves to underline the central role played by the cloud radiative feedback in enhancing the Matsuno-Gill response specifically.

\begin{figure}[ht!]
\epsscale{0.5}
\plotone{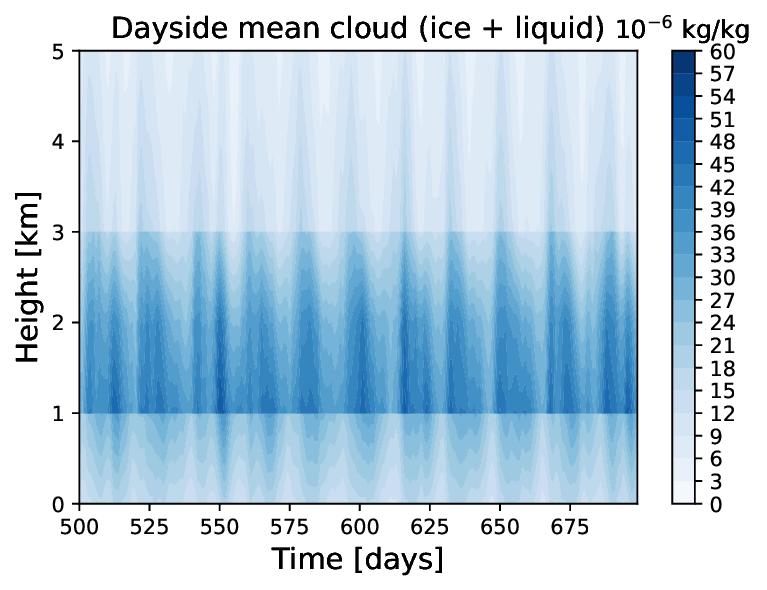}
\caption{Vertical profile of dayside mean total cloud cover (sum of ice and liquid) over time for the Control TRAP-1e simulation. As in Figure \ref{fig:resonance} a) and b), the short vertical range of 0 to 5~km results in discontinuities between vertical levels due to the low resolution of the simulation.}
\label{fig:cloudresonance}
\end{figure}

Regular 20-day cycles are visible in all quantities in the moist atmosphere case, but are absent in the dry atmosphere. The air temperature, vertical wind, and shortwave surface heating increases precede the increase in the Rossby wave power. For example, Figure \ref{fig:resonance} a), c), and g) show a peak in these three quantities at around 510 days, while the spike in the 1-0 Rossby wave (red line in Figure \ref{fig:resonance} g)) and increase in the zonal wind speed occur at 518-520 days. This pattern repeats ten times over the period displayed in the plots. Figure \ref{fig:cloudresonance} further shows variability in the dayside mean total (sum of ice and liquid) cloud cover on the same 20-day cycle. The cloud mass fraction grows during the heating/rising and drops during the cooling/subsiding part of the cycle. 

We posit an internal feedback between the dayside cloud cover and the intensity of the Matsuno-Gill response. A decrease in cloud cover allows more shortwave radiation to reach the surface, leading to atmospheric heating and subsequent ascending motion of the air mass. The 1-0 Rossby wave responds to the increase in forcing, boosting its power spectral density relative to the other constituent Rossby waves in the wind field. At the same time, the zonal wind speed increases, shifting the 1-0 wave structure further eastwards. To support this interpretation, we show in Figure \ref{fig:resonance} g) and h) both the PSD of the 1-0 wave and the summed PSD of a number of other large-scale waves, namely the 1-1, 2-1, 2-2, and 3-2 waves. At the troughs of the cycle, the PSD of the 1-0 wave is roughly equal to that of the other waves combined, and occasionally even drops below it. At the peaks, however, the PSD of the 1-0 wave increases substantially more than that of the sum of the remaining waves, indicating that this wave disproportionately receives energy during the cycle, as would be expected from its direct relationship to the Matsuno-Gill periodic forcing pattern. (Note that in the dry atmosphere case, the PSD of all waves is an order of magnitude smaller than in the moist case despite the larger shortwave flux and atmospheric temperature anomalies, highlighting the important role of moisture.)

When cloud cover increases again, less radiation reaches the surface, the air mass cools and subsides, the zonal wind speed slows, and the 1-0 Rossby wave becomes weaker and is shifted westwards from its equilibrium position. It is possible that the location of the Rossby gyres in turn affects the cloud cover, closing the causal loop, but we believe it is unlikely that the Rossby waves are the only or main factor in the density of the clouds. The zonal wind speed, which influences both the Rossby wave phase velocity and the stability of the dayside cloud cover, is also affected by the changes in the thermodynamic properties of the dayside atmosphere shown above (Figure \ref{fig:resonance} e)). Untangling these intricate relationships requires a better understanding of the factors controlling the zonal wind speed on tidally locked planets than is currently available. In addition, the cloud layer is likely to be sensitive to multiple processes in the atmosphere in addition to the zonal wind variation, including the intensity of convection, specific humidity, and the advective and radiative time scales.

\begin{table}
\hspace*{-2cm} 
\begin{tabular}{lllll}
\hline
Quantity & Control ProxB & Warm ProxB & Control TRAP-1e & Warm TRAP-1e  \\
\hline \hline
Rotation period (days) & 11.2 & 9.2 & 6.1 & 4.2 \\
Oscillation period (days) & 157.5 & 120 & 19.4 & 16.2 \\
Mean global zonal mean zonal wind (m/s) & 5.3 & 4.1 & 17.8 & 15.1 \\
Mean equator-pole temperature gradient (K/m) & -2.55$\times$10$^{-6}$ & -2.761$\times$10$^{-6}$ & -3.92$\times$10$^{-6}$ & -3.76$\times$10$^{-6}$ \\ 
\hline
\end{tabular}
\caption{Rotation period, period of Rossby gyre oscillation, mean zonal mean zonal wind, and mean meridional temperature gradient for each of the four moist atmosphere cases. The periodicity was determined from the cloud cover oscillation shown in Figure \ref{fig:cloudvar}. The meaning period for the bottom two quantities was chosen to be the same as in Figure \ref{fig:climatology}, the first 300 days of the sampling period for each simulation.}
\label{tab:period}
\end{table}

\begin{figure}[ht!]
\gridline{\fig{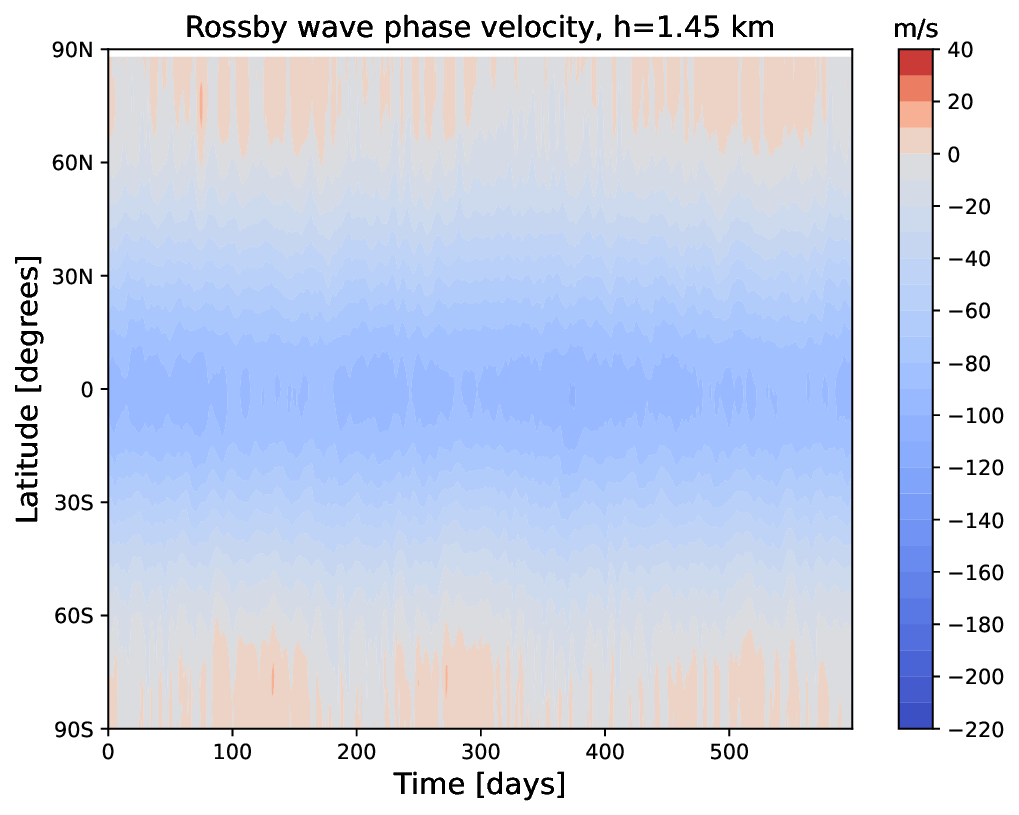}{0.45\textwidth}{a) Control ProxB}
          \fig{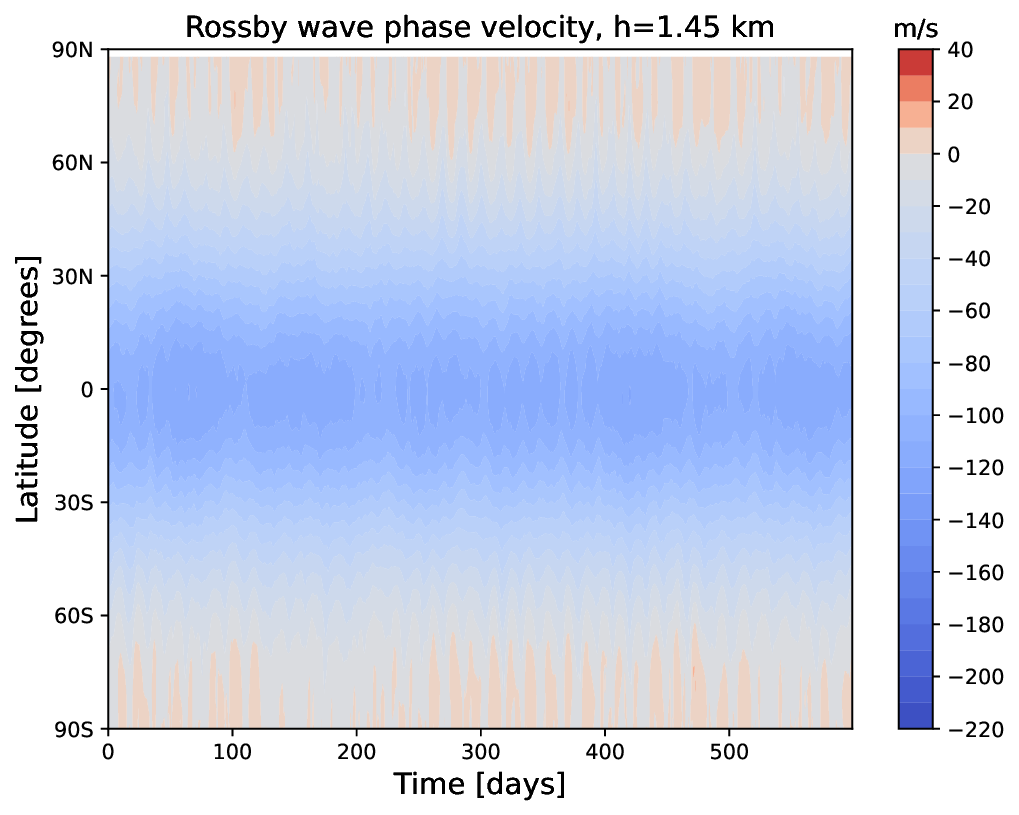}{0.45\textwidth}{b) Warm ProxB}}
\gridline{\fig{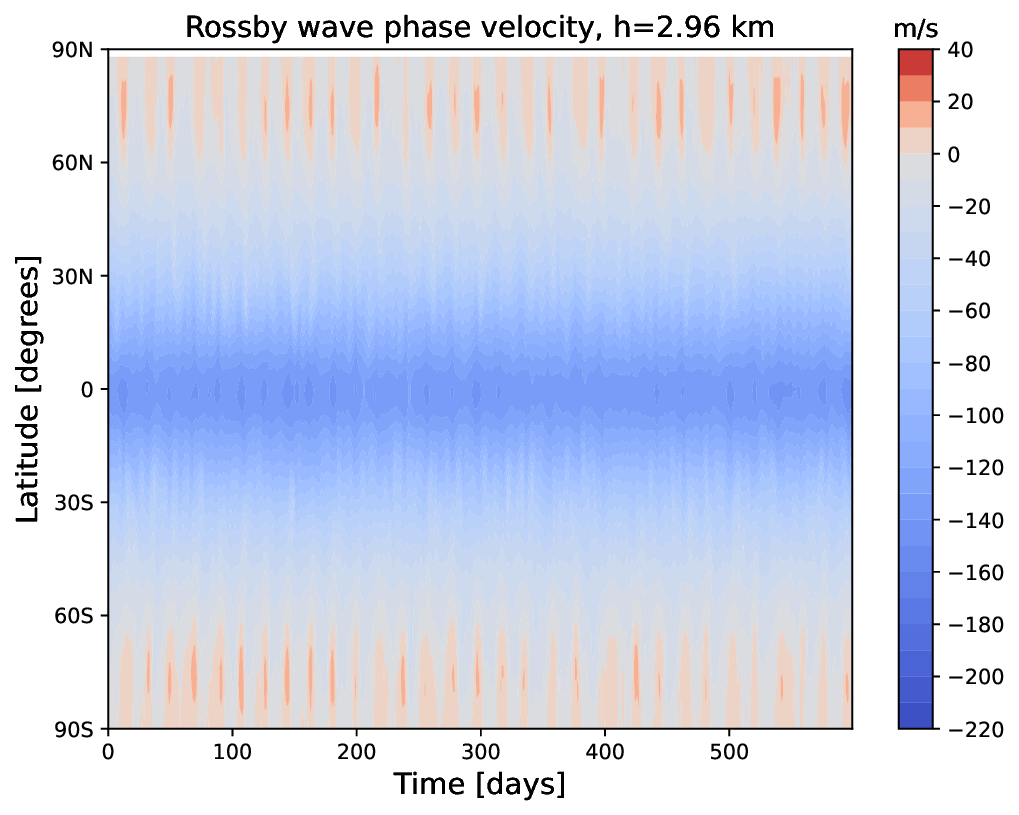}{0.45\textwidth}{c) Control TRAP-1e}
          \fig{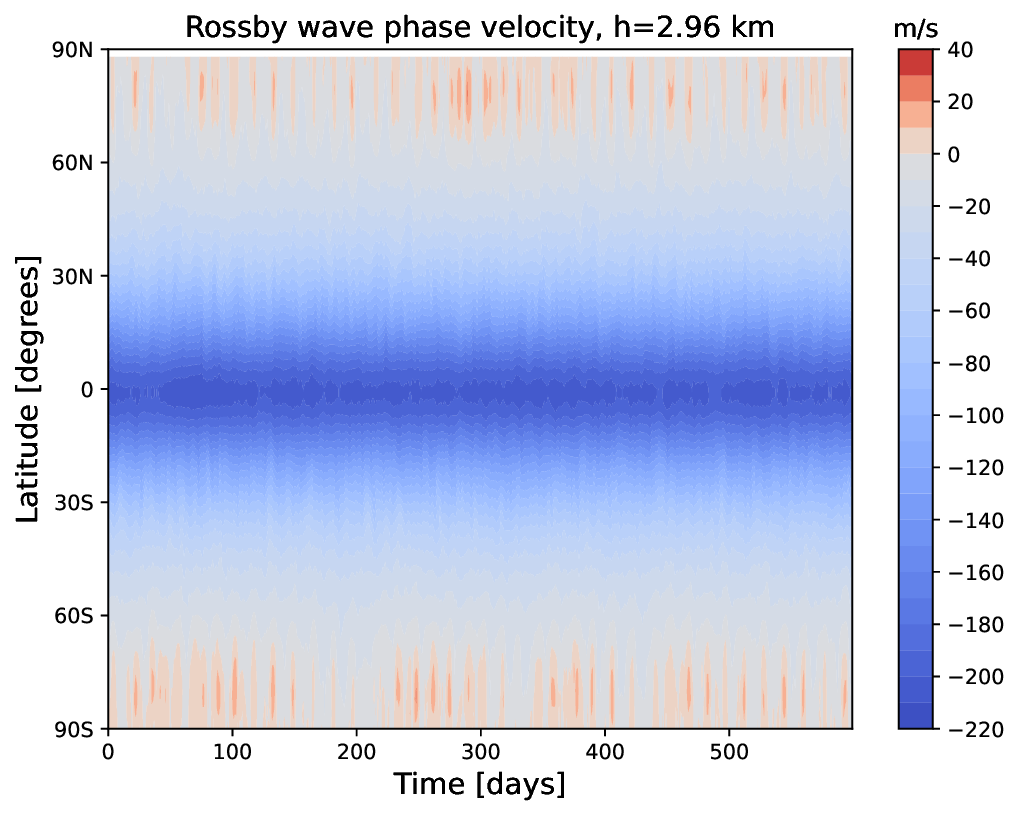}{0.45\textwidth}{d) Warm TRAP-1e}}
\gridline{\fig{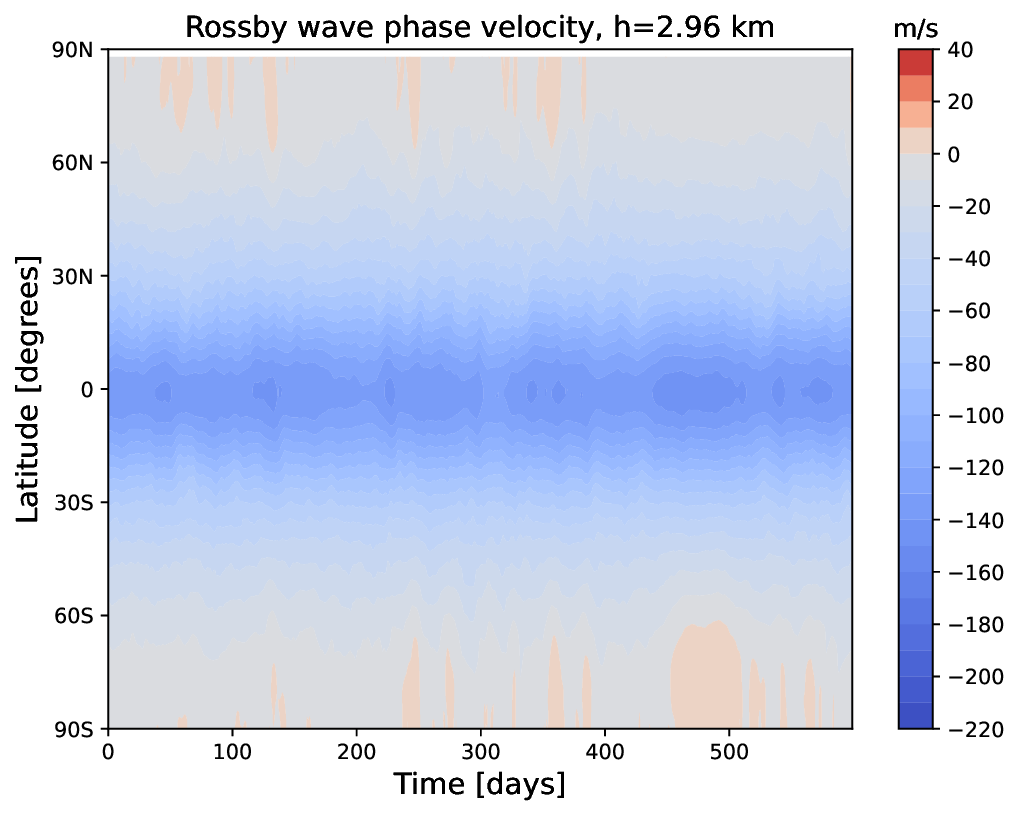}{0.45\textwidth}{e) Dry TRAP-1e}}
\caption{Latitude-time diagrams of Rossby wave phase velocity for each simulation at h=2.96 km height. Positive values correspond to eastward flow, while negative values represent westward flow. The phase velocity is calculated with subtraction of the mean zonal wind as in Figure \ref{fig:rossbycp}.}
\label{fig:cphase}
\end{figure}

The period of the oscillation, given in Table \ref{tab:period}, varies substantially between the four moist atmosphere cases. Understanding why the period is longer in some simulations is important because a slower oscillation implies the planet will have a longer period of clear skies at the limb, potentially allowing for repeat observations when conditions are favorable. We find that the oscillation period monotonically decreases in parallel with the rotation period, but the relationship is not linear. We expect the rotation period to influence the Rossby wave phase velocity directly through $\beta$. However, other factors are clearly in play. While the rotation period decreases by similar amounts (2-3 days) between each simulation, there is a disproportionately large difference between the oscillation periods of the Proxima Centauri b and TRAPPIST-1e simulations. We believe the non-linearity can be explained by the additional influence of the zonal wind on the Rossby wave phase velocity as defined in Equation \ref{eqn:cp}. Figure \ref{fig:cphase} shows latitude-time diagrams of the Rossby wave phase velocity for each simulation. Both the eastward and westward phase of the oscillation display higher phase velocities in the TRAPPIST-1e simulations as compared to Proxima Centauri b, accounting for the much shorter oscillation periods of the former. Effectively, the more rapid background zonal flow is shifting the wave response east and west of its equilibrium point more quickly in the TRAPPIST-1e cases.

The phase velocities differ because the zonal mean wind, reported in Table \ref{tab:period}, jumps significantly between the Proxima Centauri b (4-5 m/s) and TRAPPIST-1e (15-18 m/s) simulations. In particular, Figure \ref{fig:climatology} shows that the zonal mean wind speeds are higher at the mid-latitudes where the gyres are centered in the TRAPPIST-1e cases than the Proxima Centauri b cases. This is because Control and Warm TRAP-1e are both in the double mid-latitude jet circulation regime described by \cite{sergeev_bistability_2022}, while Control and Warm ProxB are both in the single equatorial jet regime. \cite{sergeev_bistability_2022} found that the single jet regime is characterized by transport of angular momentum to the equator by the stationary eddy term of the axial angular momentum equation driven by wave-jet resonance between the equatorial jet and the zonal wavenumber-1 Rossby wave (consistent with the findings of e.g., \cite{wang_phase_2021}, \cite{hammond_wave-mean_2018}, and \cite{tsai_three-dimensional_2014}). The double jet regime, on the other hand, forms when the mean advection and transient eddy terms of the momentum budget become large at the mid-latitudes and the stationary eddy term decreases at the equator: the mid-latitude jets speed up while the equatorial jet slows down. As the stationary gyres form in the mid-latitudes, it is the wind speeds in the mid-latitudes that control the magnitude of the Rossby wave phase velocity.

\cite{sergeev_bistability_2022} explores the development of both the single jet and the double jet regimes during the model spin-up period. As shown in Table \ref{tab:period}, the TRAPPIST-1e simulations have a larger equator-to-pole temperature gradient than the Proxima Centauri b simulations. This larger temperature gradient may cause greater baroclinic instability, promoting the formation of baroclinic mid-latitude jets. However, the exact reasons why a simulated planet is nudged into one regime or the other may be varied and are difficult to isolate in a model as complex as the UM. According to the weak temperature gradient theory applicable to slowly rotating tidally locked planets \citep{pierrehumbert_atmospheric_2019}, the equator-pole temperature gradient should increase with increasing rotation rate, which is consistent with the values in Table \ref{tab:period}. The discontinuity between the two Proxima Centauri b and the two TRAPPIST-1e simulations is a reflection of the shift from the single jet to the double jet regime. These patterns broadly suggest that more slowly rotating planets with weaker equator-pole temperature gradients and a single equatorial jet are likely to have longer oscillation periods and longer windows of cloudless sky at the terminators.

\subsection{Cloud variability and observables} \label{subsec:impacts}
\begin{figure}[ht!]
\gridline{\fig{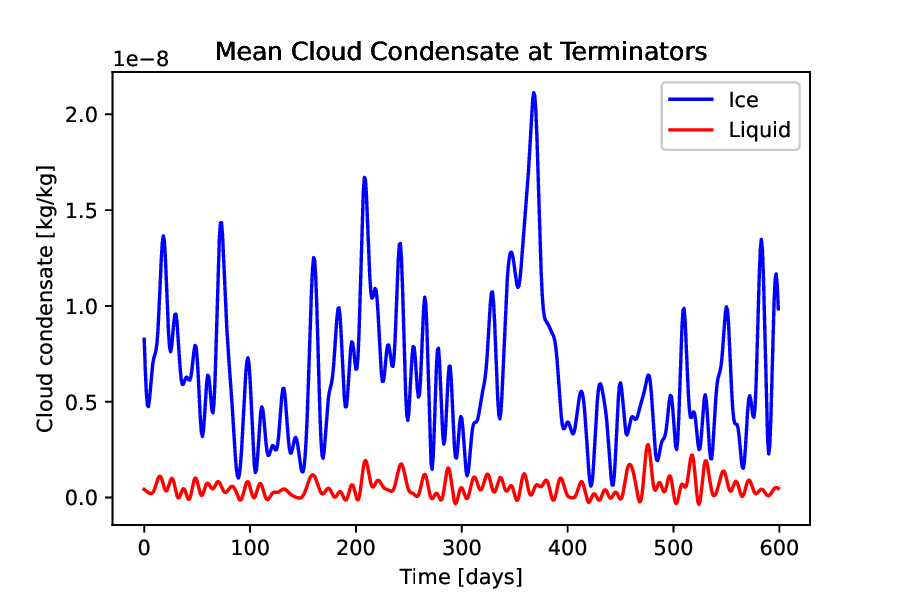}{0.5\textwidth}{a) Control ProxB}
          \fig{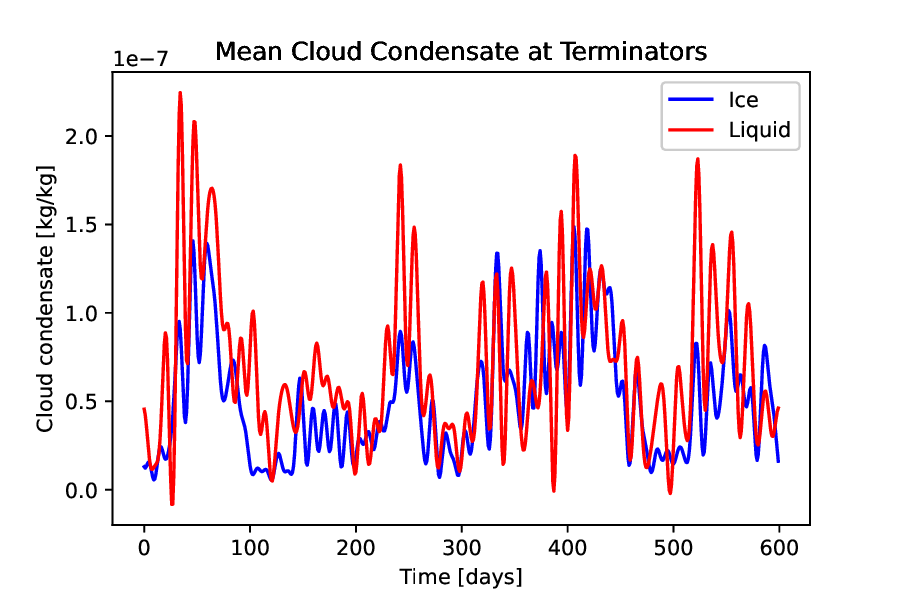}{0.5\textwidth}{b) Warm ProxB}}
\gridline{\fig{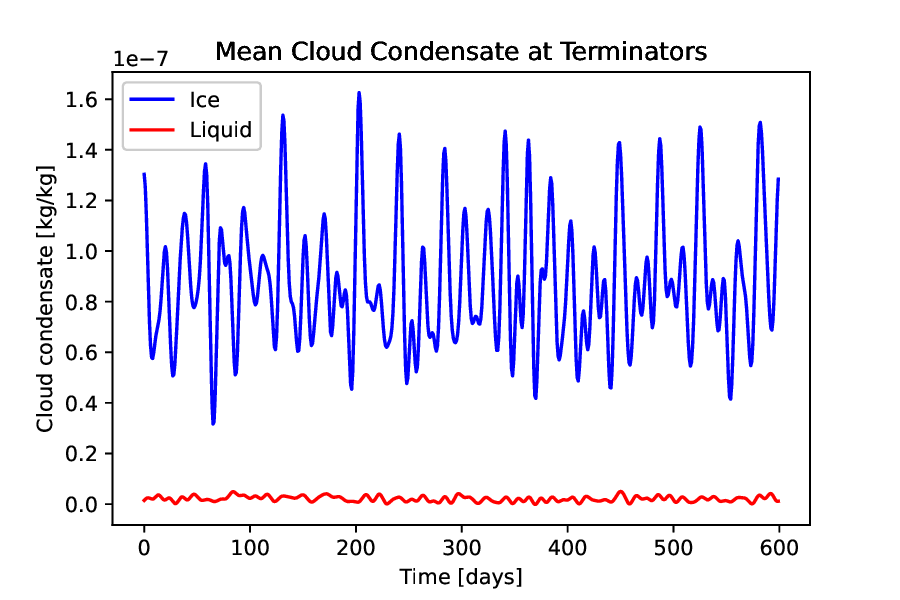}{0.5\textwidth}{c) Control TRAP1-e}
         \fig{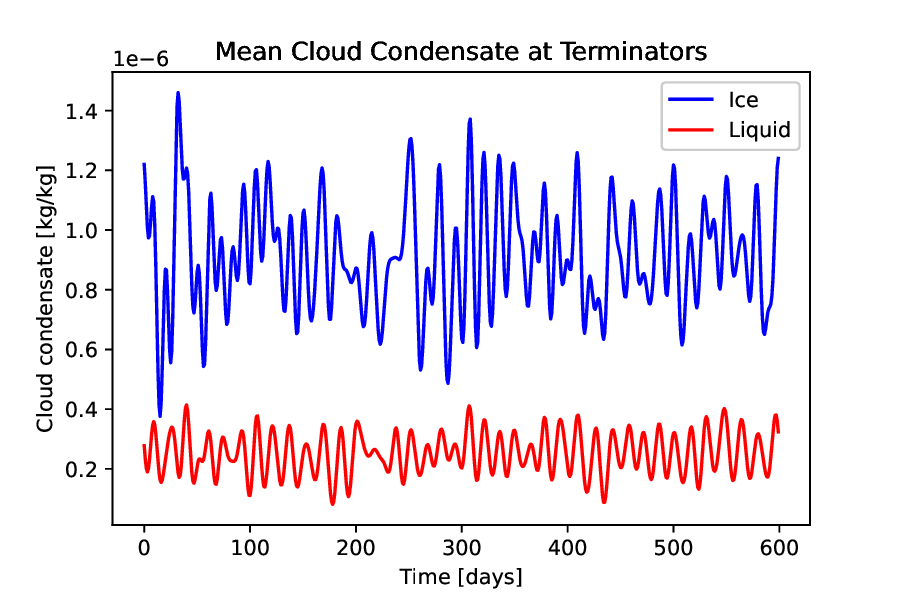}{0.5\textwidth}{d) Warm TRAP1-e}}
\caption{Mean cloud condensate (mixing ratio, kg/kg) over time at the planetary limb for each of the four moist atmosphere simulations. Liquid and ice cloud are shown separately. Each type of cloud is averaged over all latitudes and all heights on the eastern and western terminator. The data has been filtered to remove cycles with periods shorter than 10 days. Note the different limits of the y-axis.}
\label{fig:cloudvar}
\end{figure}

The migration of Rossby gyres impacts the amount of moisture transport and thus cloud condensate at the planetary terminators. All four of the moist atmosphere simulations display cloud cover variability at the terminators, shown in Figure \ref{fig:cloudvar}. The Control and Warm ProxB runs in Figure \ref{fig:cloudvar} a) and b), respectively, exhibit large fluctuations in cloud condensate in the observable regions of the planet, ranging from near 0 to $7\times10^{-7}$ kg/kg and $2.5\times10^{-6}$ kg/kg, respectively, on a time scale matching the migratory cycle of the Rossby gyres (120 days/160 days). The TRAPPIST-1e simulations do not undergo these long-period cycles, but show regular smaller magnitude fluctuations on an approximately 20-day time scale.

\begin{figure}[ht!]
\gridline{\fig{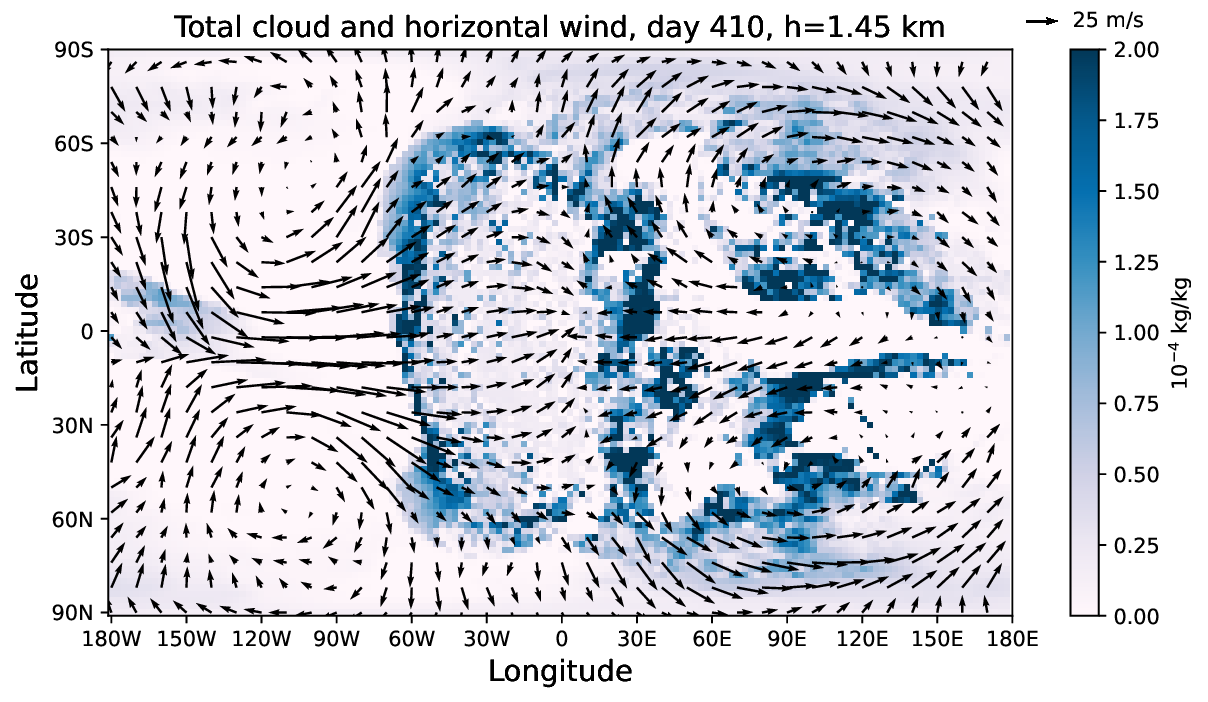}{0.5\textwidth}{a) Warm ProxB, day 410}
          \fig{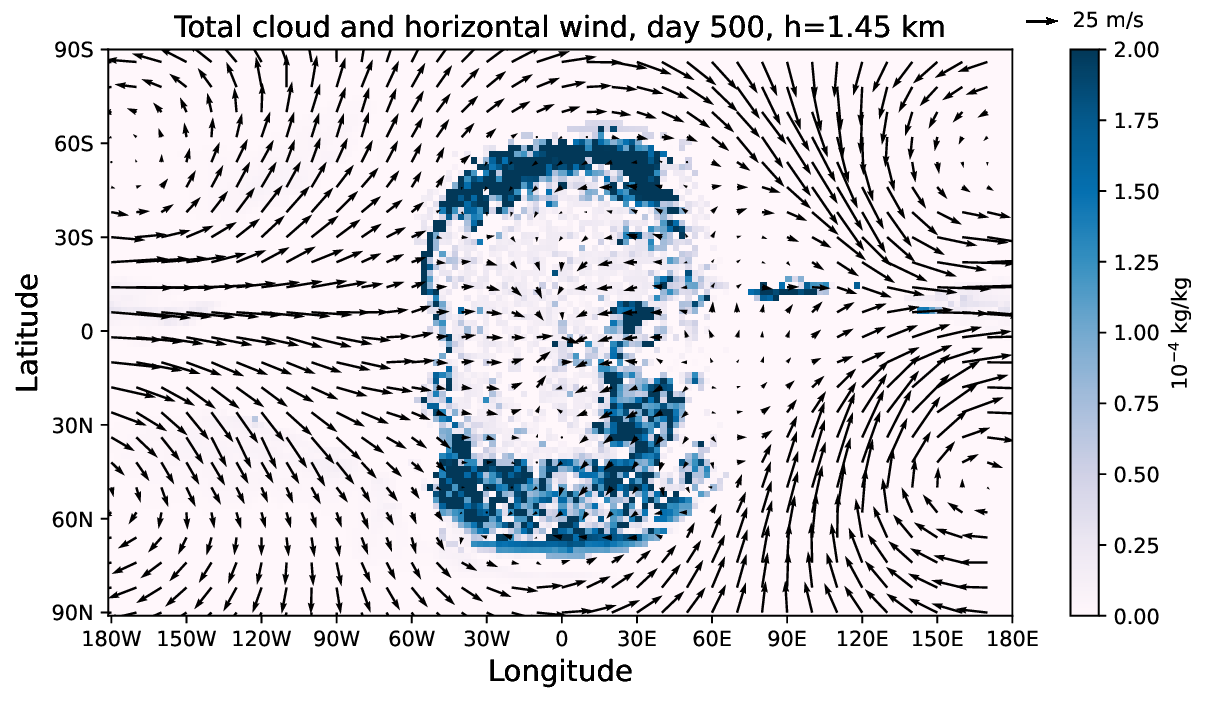}{0.5\textwidth}{b) Warm ProxB, day 500}}
\gridline{\fig{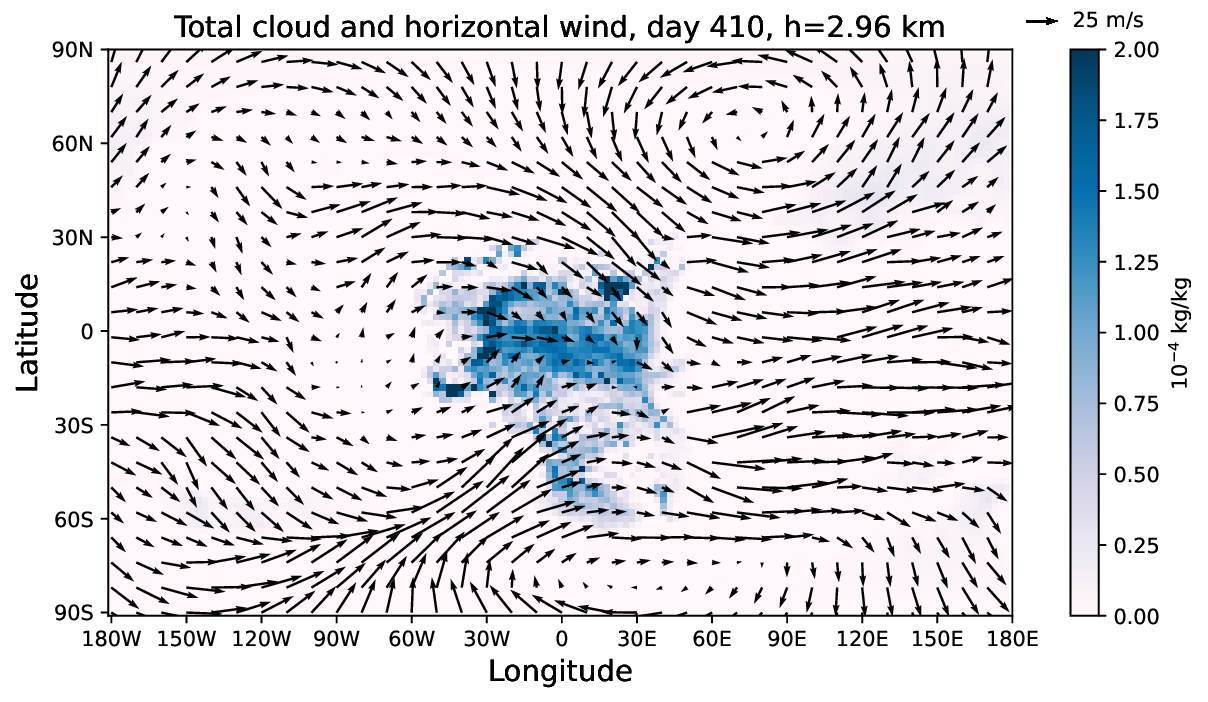}{0.5\textwidth}{c) Control TRAP-1e, day 410}
         \fig{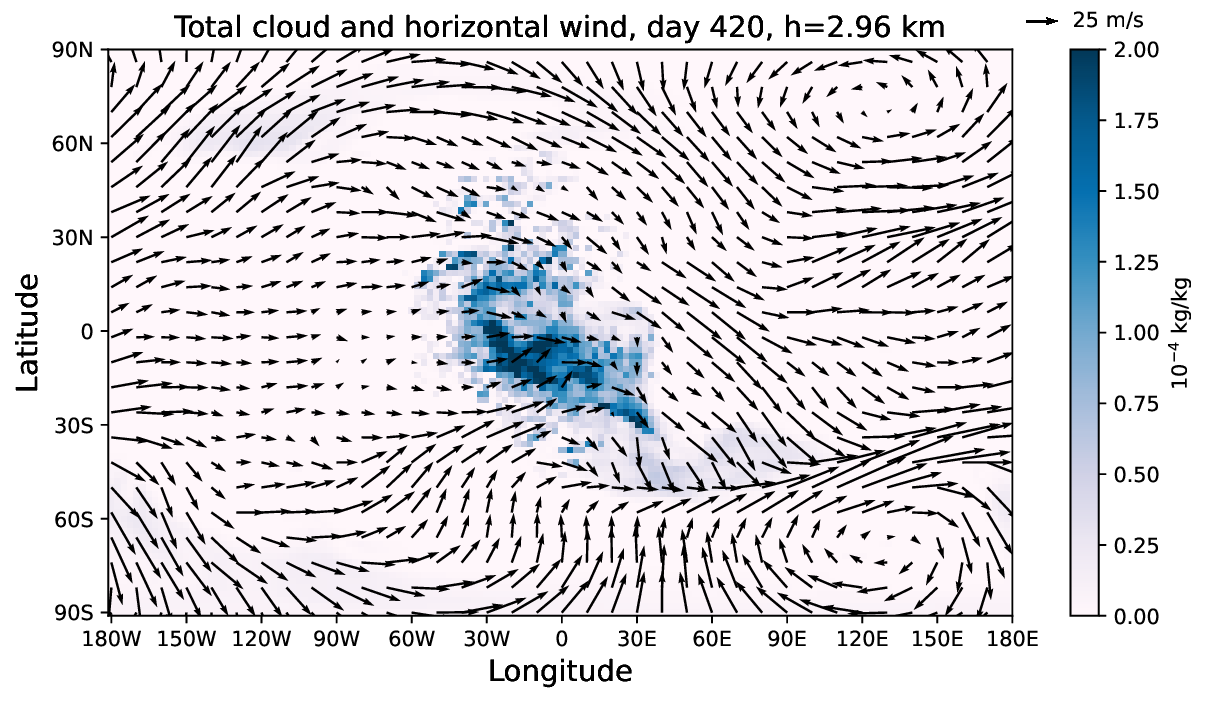}{0.5\textwidth}{d) Control TRAP-1e, day 420}}
\caption{Wind vectors overlaid on the total cloud condensate (sum of ice and liquid) at the given height for Warm ProxB and Control TRAP-1e simulations. The images are daily snapshots chosen to illustrate the eastmost and westmost phases of the Rossby gyre migration for each planet.}
\label{fig:cloudquiver}
\end{figure}

Figure \ref{fig:cloudquiver} depicts the interaction between the wind field and the dayside clouds. Figure \ref{fig:cloudquiver} a) and b) are two stages in the Rossby gyre migratory cycle for the Warm ProxB simulation, corresponding to a cloud condensate maximum and minimum. During the maximum, the eastern pair of Rossby gyres is at the extreme western part of its propagation path, where it intersects with the region of heavy cloud cover around the substellar point. During the minimum, the gyres are at the extreme eastern part of the propagation path and do not interact with the dayside clouds. In the TRAPPIST-1e case, shown in Figure \ref{fig:cloudquiver} c) and d), the Rossby gyres are too far polewards to interact with the dayside cloud cover. The short-period cycles shown in Figure \ref{fig:cloudvar} c) are likely a direct reflection of the fluctuation in cloud condensate and moisture described in \ref{subsec:mechanism} and shown in Figures \ref{fig:resonance} and \ref{fig:cloudresonance}, on which the longer-period effect from the traveling wave structures is overlaid.

\begin{figure}[ht!]
\gridline{\fig{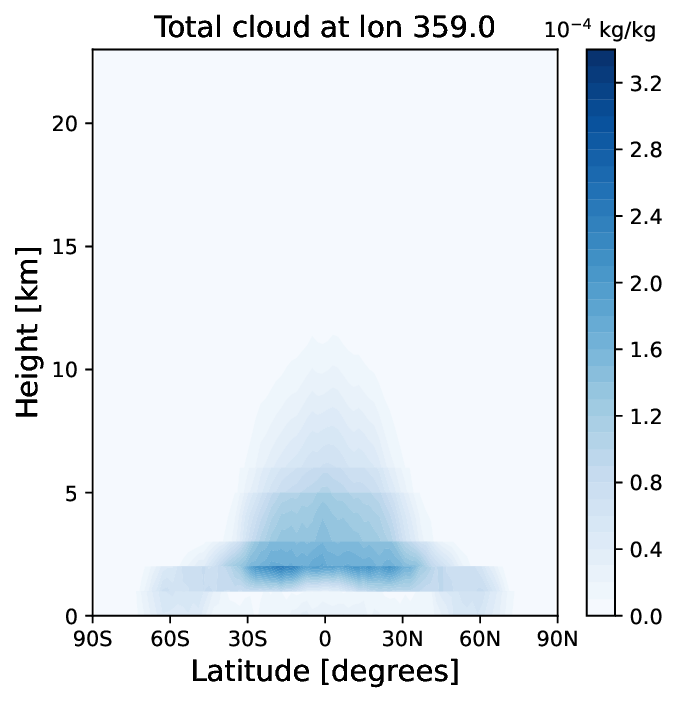}{0.2\textwidth}{a) Control ProxB}
         \fig{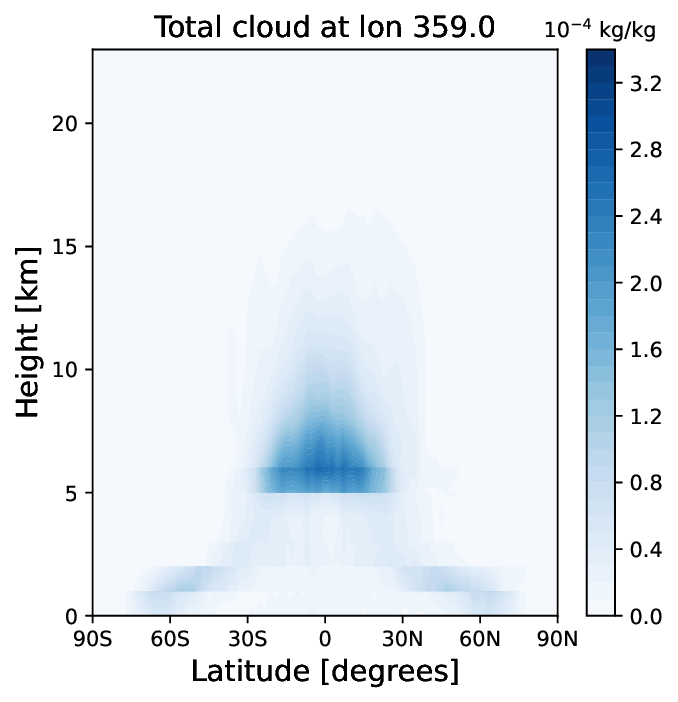}{0.2\textwidth}{b) Warm ProxB}
         \fig{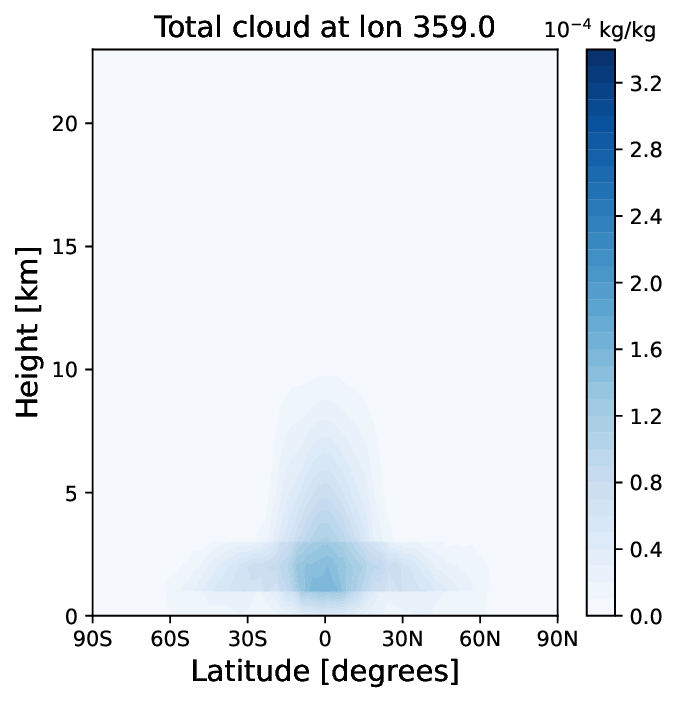}{0.2\textwidth}{c) Control TRAP-1e}
         \fig{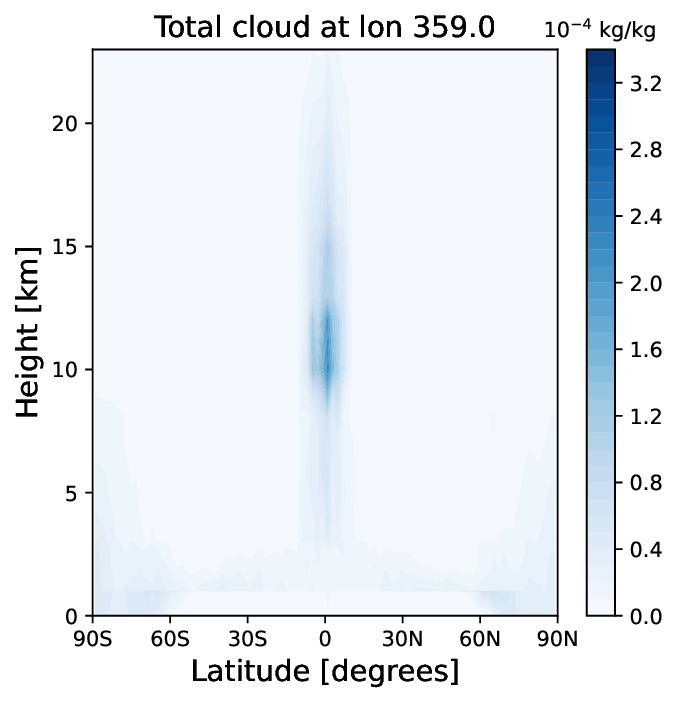}{0.2\textwidth}{d) Warm TRAP-1e}}
\gridline{\fig{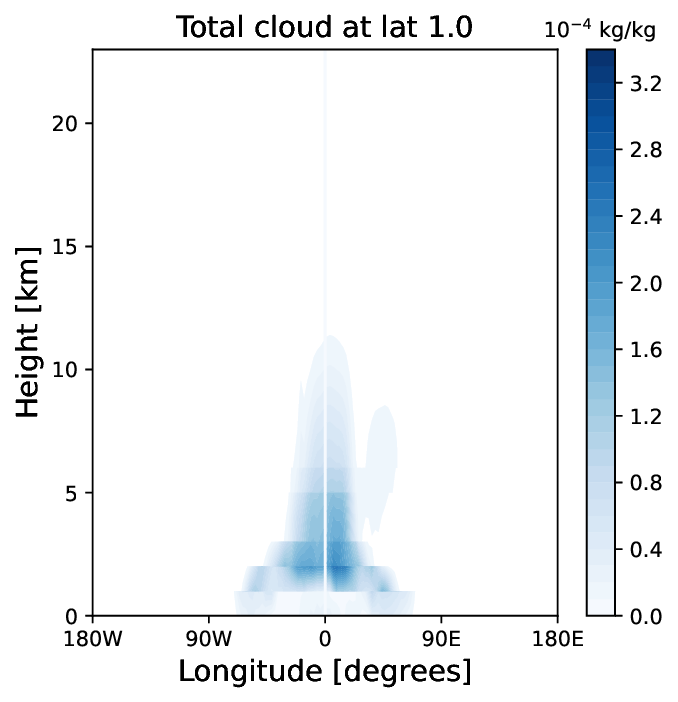}{0.2\textwidth}{e) Control ProxB}
         \fig{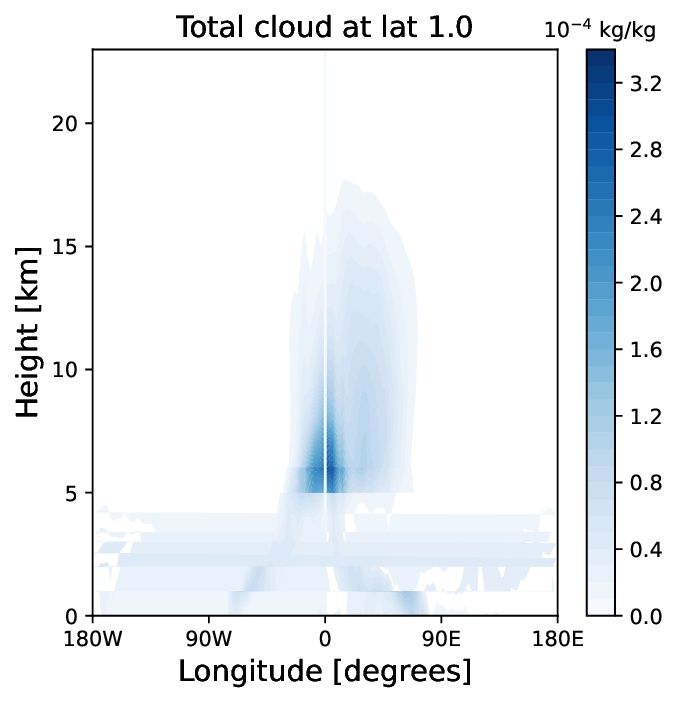}{0.2\textwidth}{f) Warm ProxB}
         \fig{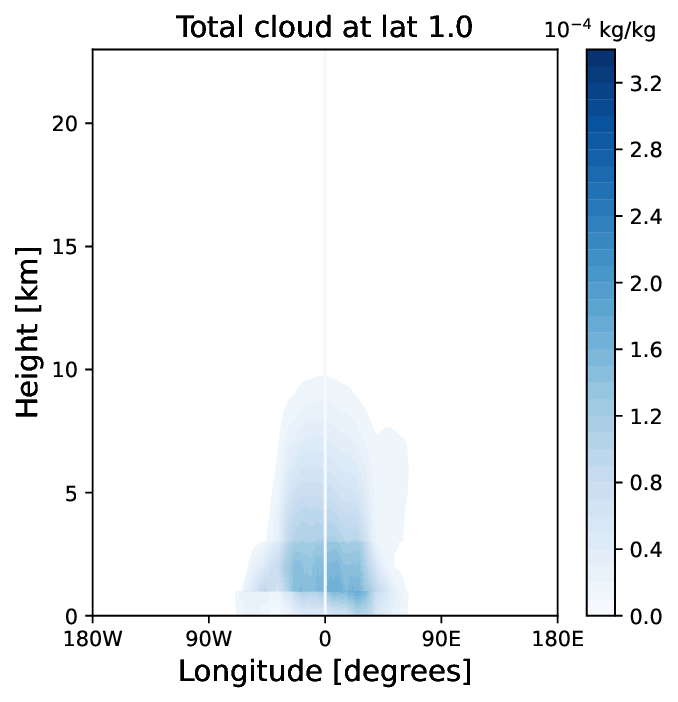}{0.2\textwidth}{g) Control TRAP-1e}
         \fig{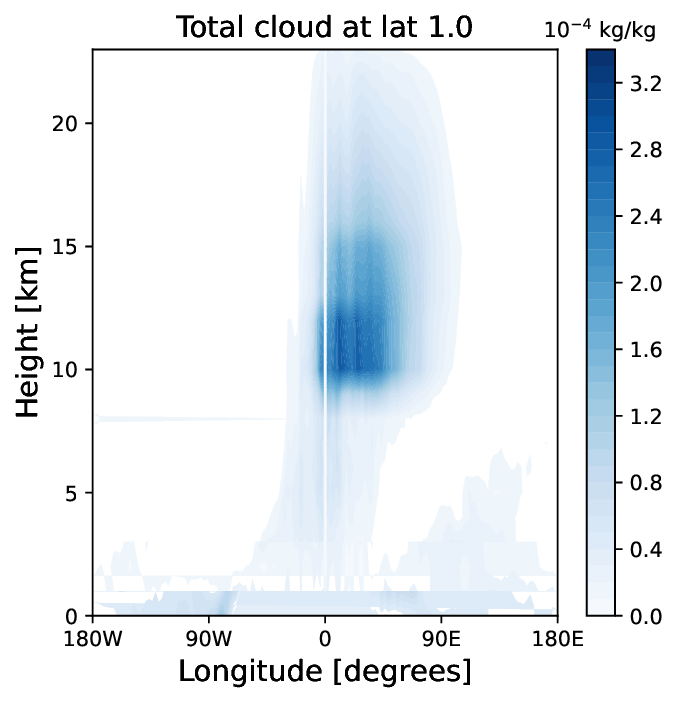}{0.2\textwidth}{h) Warm TRAP-1e}}         
\caption{Longitudinal and latitudinal cross-sections of the dayside cloud layer for the four moist atmosphere simulations. The total cloud is the sum of ice and liquid cloud condensate. The images depict 120-day means.}
\label{fig:cloudbubbles}
\end{figure}

The magnitude and periodicity of the variation in cloud cover at the planetary limb is highly sensitive to not only the Rossby wave propagation, but also the dayside cloud structure. As demonstrated by the TRAPPIST-1e cases, the amount of cloud at the terminator will not be affected by the Rossby wave oscillation unless the Rossby gyres form at low or mid-latitudes where they can advect cloud from the substellar region. Figure \ref{fig:cloudbubbles} shows cross-sections of the dayside cloud layer at the equator and at longitude 0 for the four moist atmosphere simulations. The extent of the cloud cover in longitude, latitude, and altitude depends on the temperature and moisture profile of each simulated planet, but as the longitude, latitude, and even peak altitude of the Rossby waves also vary in different simulations, the parameter space of the resulting wave-cloud interaction is complex.  

\begin{figure}[ht!]
\gridline{\fig{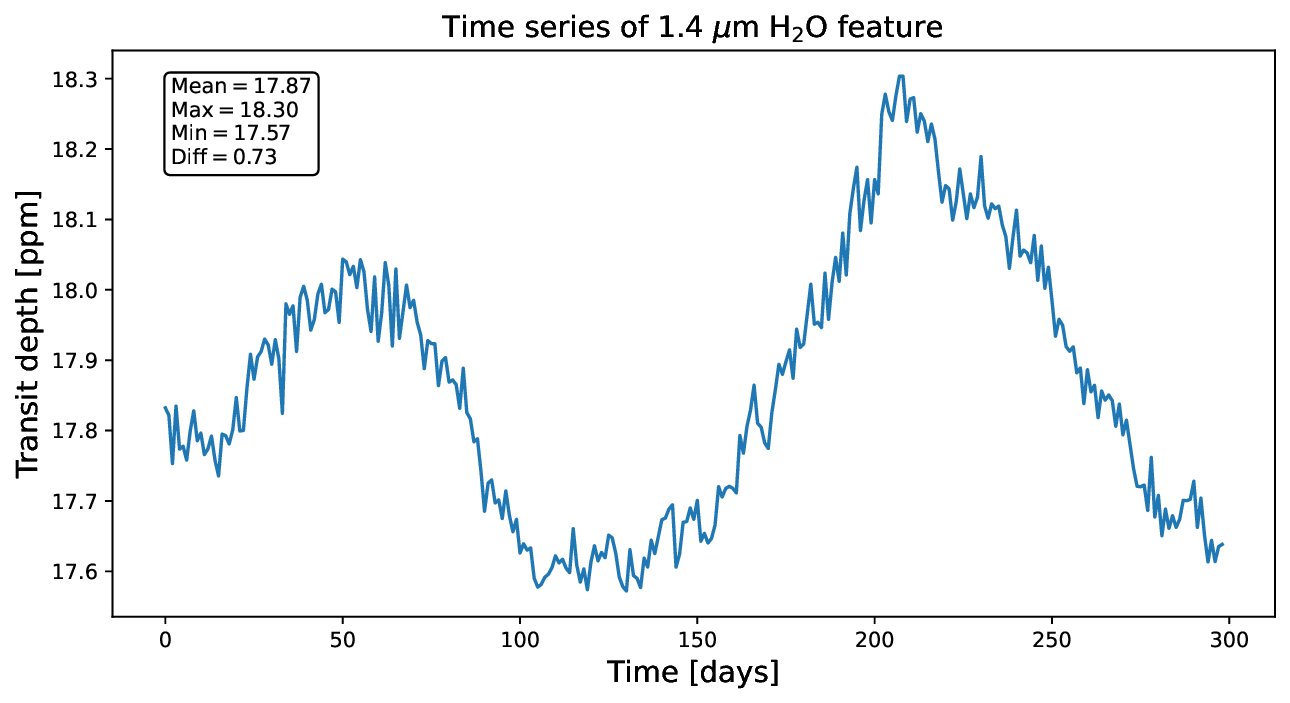}{0.5\textwidth}{a) Water feature, Control ProxB}
          \fig{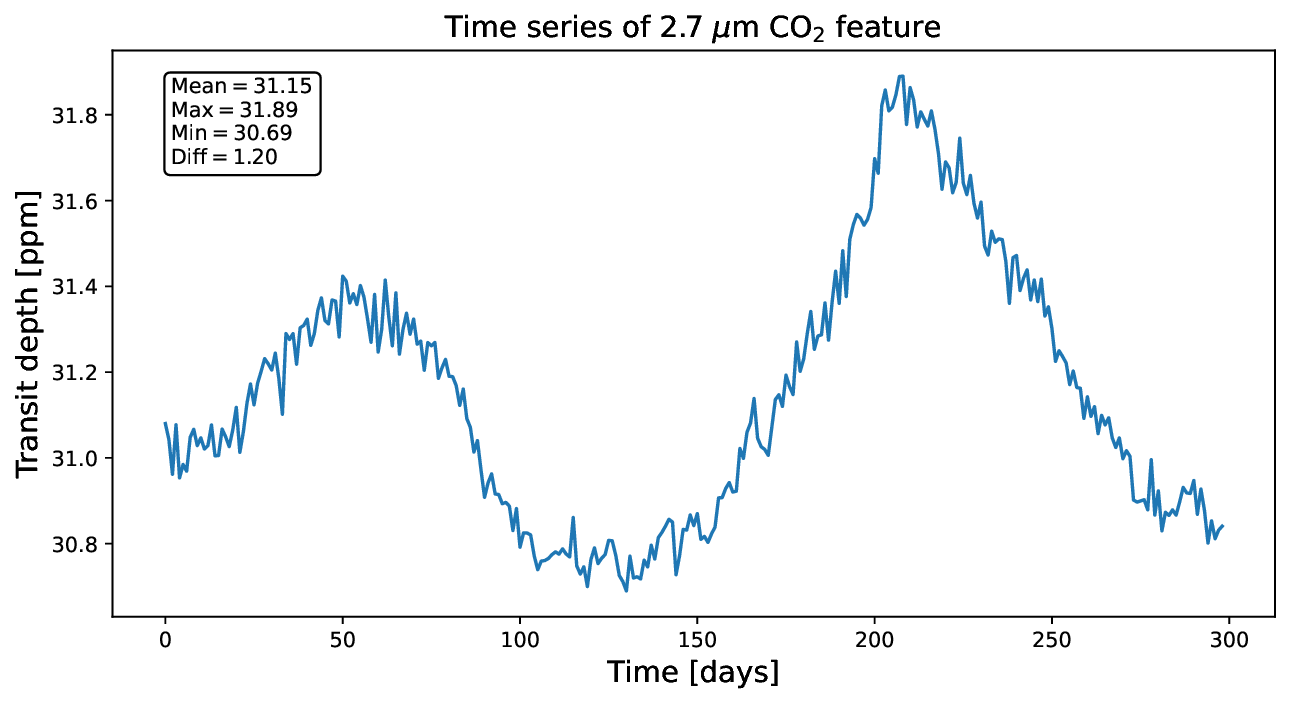}{0.5\textwidth}{b) CO$_2$ feature, Control ProxB}}
\gridline{\fig{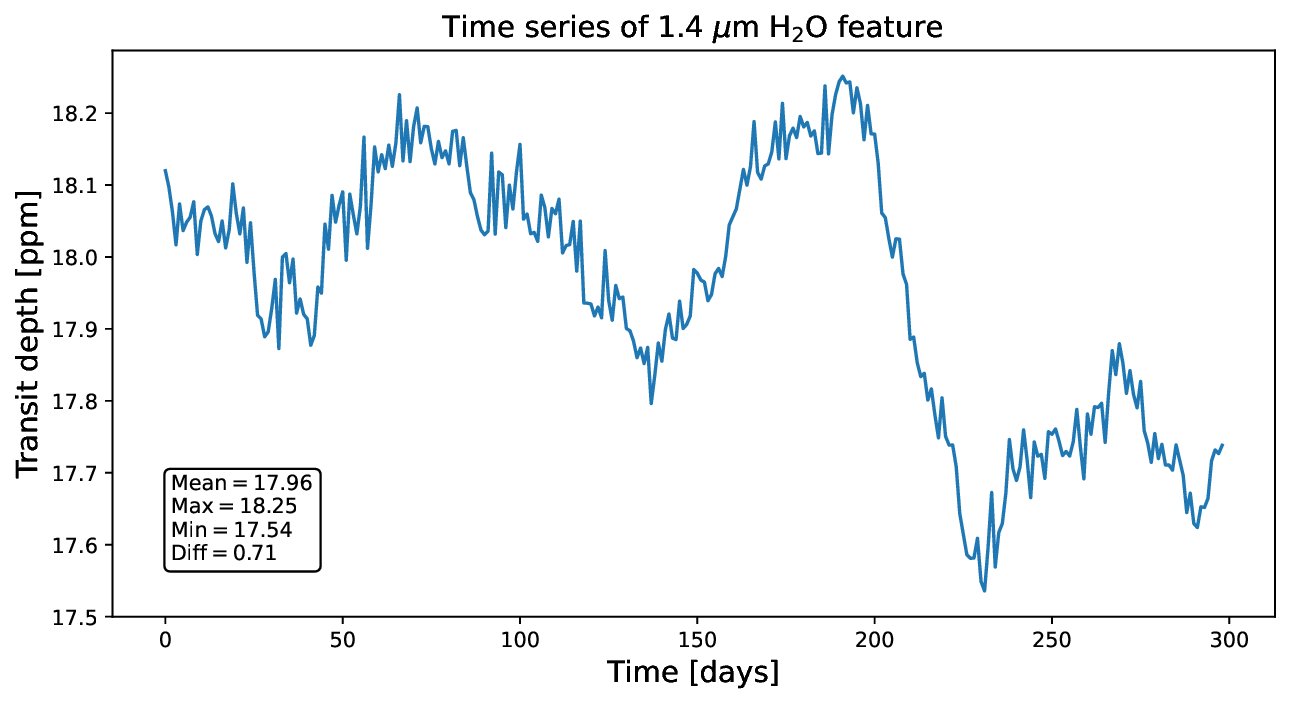}{0.5\textwidth}{c) Water feature, Warm ProxB }
          \fig{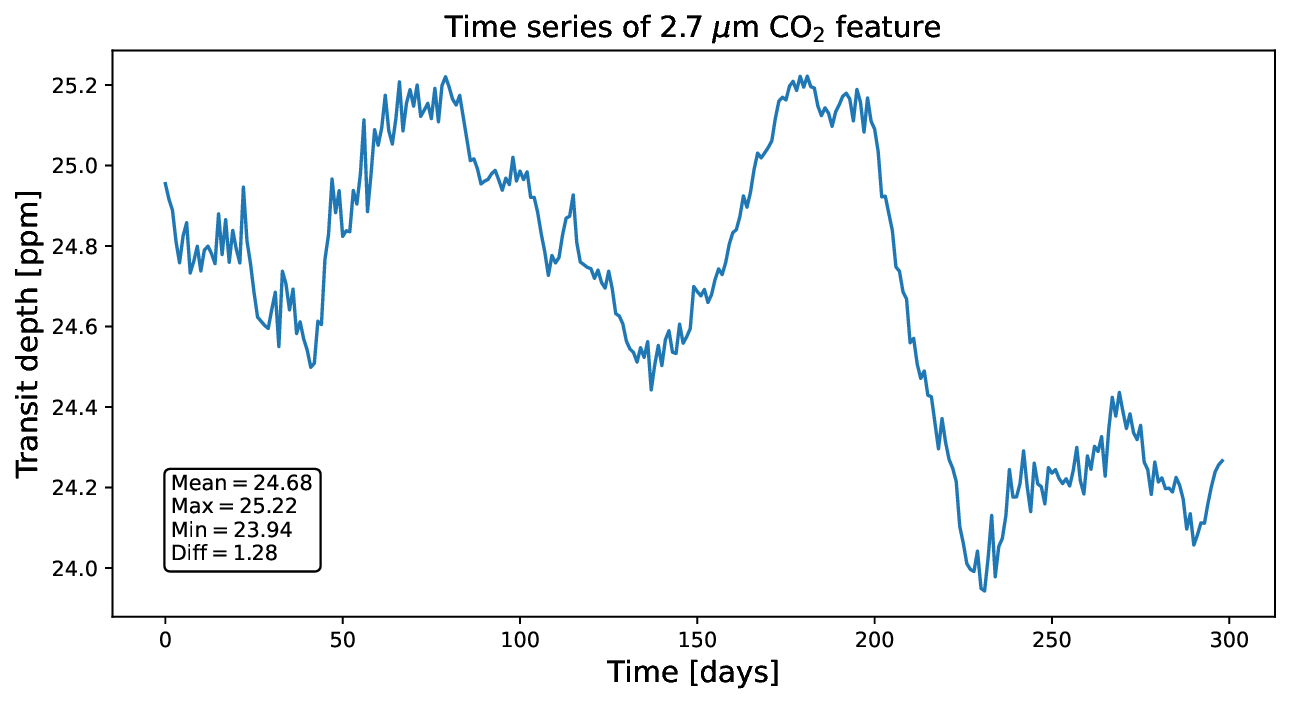}{0.5\textwidth}{d) CO$_2$ feature, Warm ProxB}}
\gridline{\fig{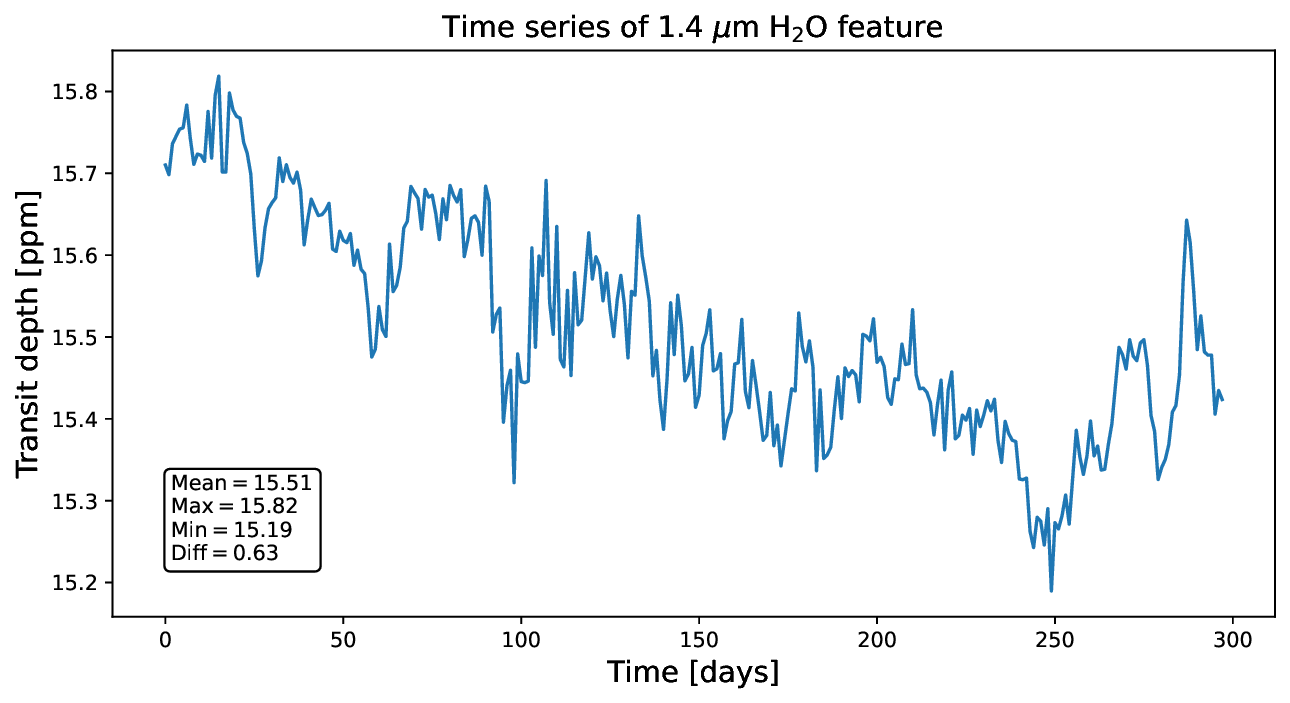}{0.5\textwidth}{e) Water feature, Control TRAP-1e}
          \fig{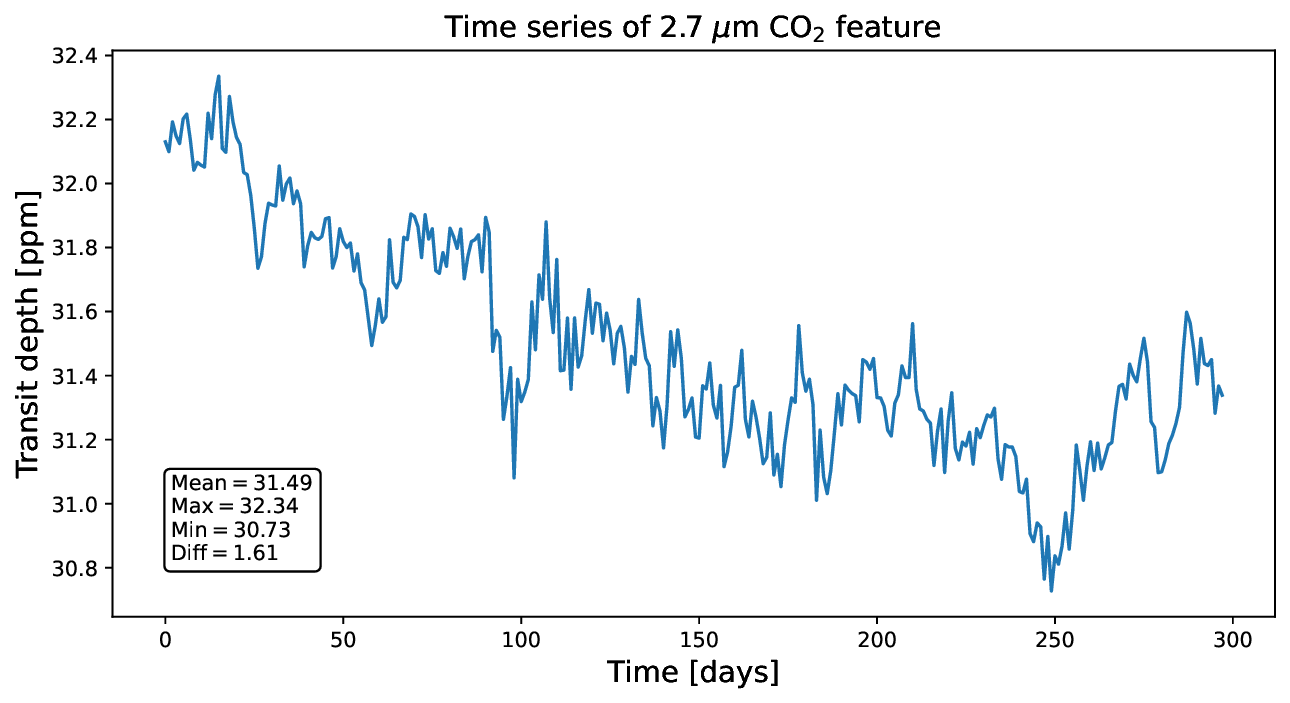}{0.5\textwidth}{f) CO$_2$ feature, Control TRAP-1e}}
\gridline{\fig{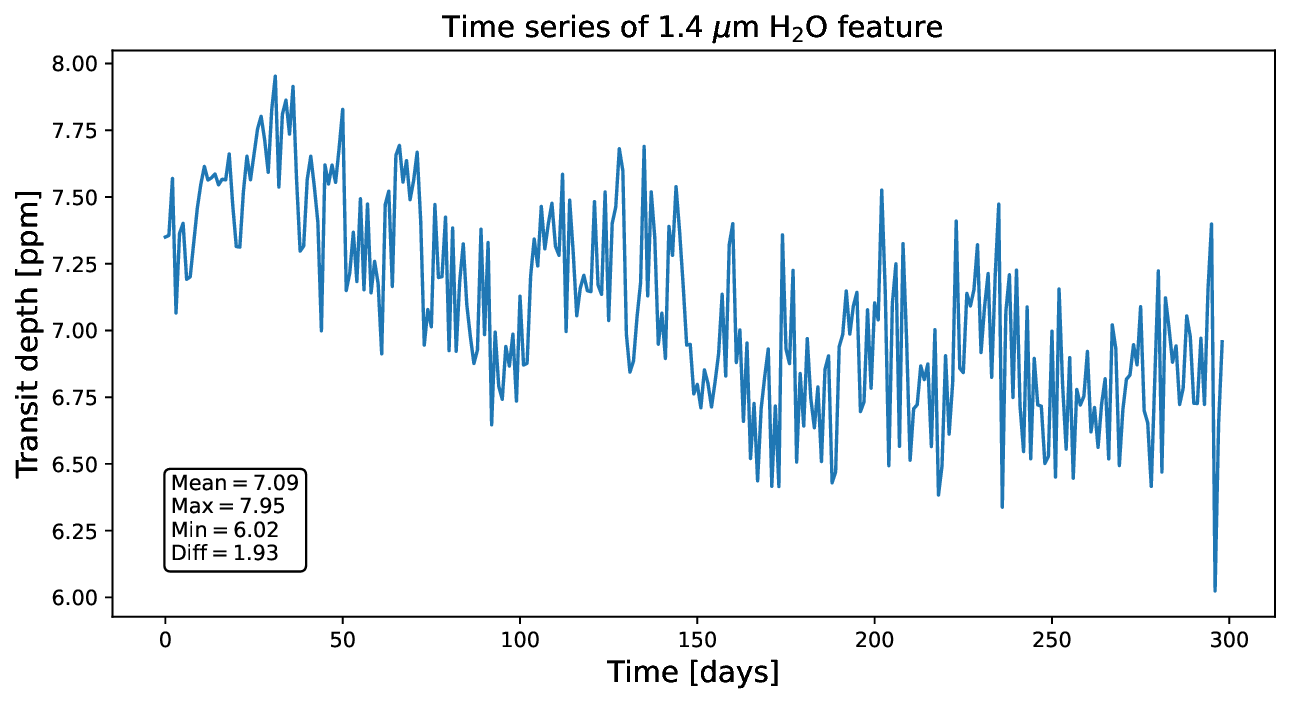}{0.5\textwidth}{g) Water feature, Warm TRAP-1e}
          \fig{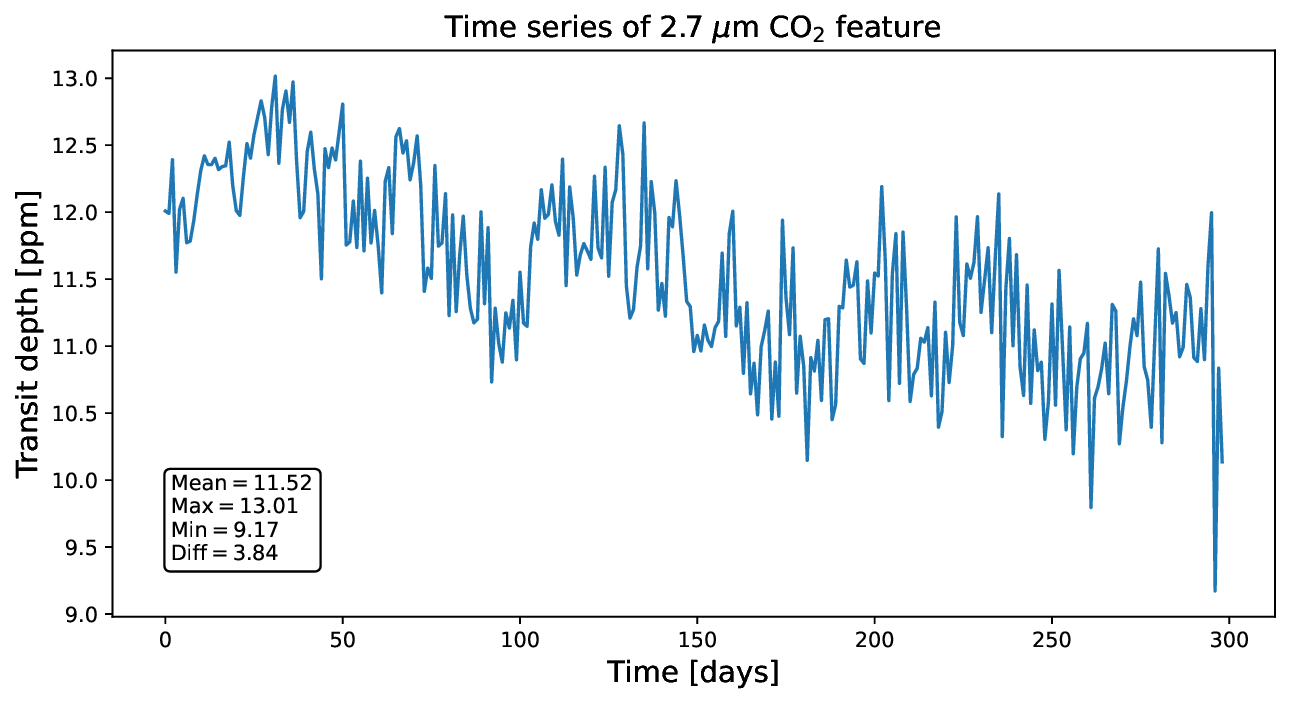}{0.5\textwidth}{h) CO$_2$ feature, Warm TRAP-1e}}
\caption{Time series of the 1.4 $\mu$m water absorption feature and 2.7 $\mu$m CO$_2$ absorption feature for the Control ProxB, Warm ProxB, Control TRAP-1e, and Warm TRAP-1e simulations. A sample of 300 days is taken to cover the 157.5 and 120 day oscillation periods in the Control and Warm Proxima Centauri b simulations, respectively.}
\label{fig:spectra}
\end{figure}

To explore the potential impact of wave-cloud interactions on observations, we simulated transit spectra for the Control ProxB, Warm ProxB, Control TRAP-1e, and Warm TRAP-1e simulations, excluding the Dry TRAP-1e simulation as it does not form clouds. We constructed time series of two absorption features, shown in Figure \ref{fig:spectra}. We chose the water line at 1.4 $\mu$m because it does not overlap with any CO$_2$ features and the CO$_2$ feature at 2.7 $\mu$m because it is a strong line in the available NIRSpec spectrum and does not overlap with the N$_2$-N$_2$ collision-induced absorption at 4.3 $\mu$m. As the CO$_2$ abundance in the simulations is fixed, variability in the transit depth of this feature can only be due to differences in the muting effect of cloud cover or due to temperature fluctuations and not due to variations in CO$_2$ content. For the water feature, variations in transit depth may also be due to differences in water vapor content on different days, caused by other factors such as the LASO and random fluctuation (model noise). However, the time series for the H$_2$O and CO$_2$ are well-correlated, supporting clouds as a factor in the variability. In the Control and Warm ProxB simulations in Figure \ref{fig:spectra} a)-d), the time series show clear long-period variation in addition to small continuous fluctuations, but the relative difference in the transit depths for these simulations is only 4-5 \%.  The percentage variation for the Warm TRAP-1e simulation is the largest in the comparison at 18-20 \%, but the transit depths are profoundly muted compared to Control TRAP-1e because of the high cloud deck visible in Figure \ref{fig:cloudbubbles} d) and h). In addition, while the Control and Warm ProxB time series have extended periods of larger transit depths, corresponding to the longer period of the cloud cover oscillation for this planet, the TRAP-1e runs lack these multi-week periods of stronger transit signals due to their shorter Rossby wave cycle as described in Section \ref{subsec:mechanism}. 

\section{Discussion}\label{sec:discussion}
Our results support the existence of internal atmospheric variability on tidally locked aquaplanets even in the absence of a varying stellar spectrum or stellar activity, rotation with respect to the host star, obliquity, and eccentricity. The stabilizing feedback between the dayside cloud cover and atmospheric temperature, identified in previous work on the inner edge of the habitable zone for tidally locked Earth-like planets \citep{kopparapu_inner_2016, yang_strong_2014,yang_stabilizing_2013}, induces periodicity in the atmospheric dynamics. \cite{edson_atmospheric_2011} postulated that the large amplitude of the zonal wavenumber 1 Rossby wave on slowly rotating planets with superrotating atmospheres is caused by resonance between this wave and the spatially periodic heating. Our finding of a disproportionate increase in the power spectral density of the $1-0$ wave directly after an increase in net surface shortwave flux supports their hypothesis. 

The increase in surface heating is caused by a drop in total cloud cover in the dayside, reducing the cloud albedo. This periodic reduction in cloud cover may be affected by the propagation of the Rossby gyres in a closed feedback loop, but it is likely that other aspects of the circulation, especially the magnitude of the zonal and vertical winds and intensity of convection in the substellar region also play a role. The potential relationship between atmospheric moisture and cloud-radiative feedback is reminiscent of theories of the Madden-Julian Oscillation (MJO) \citep{zhang_four_2020, zhang_madden-julian_2005}, particularly the moisture-mode hypothesis, which also posits a planetary wave response. In the moisture-mode theory of the MJO, clouds trap longwave radiation in the troposphere, leading to enhanced areas of column moisture, convection, and precipitation. These areas are collocated with corresponding areas of dry air and suppressed convection to their east. The spatially periodic heating anomalies (which are regions of divergence) cause a planetary wave response which is the moist atmosphere analogue of the Matsuno-Gill dry atmosphere wave response. The planetary wave response in turn advects moisture horizontally eastwards, propagating the precipitation/moisture anomaly eastwards. As there is no consensus about the mechanism of the MJO, however, and the complexity of the factors influencing the cloud cover on the dayside is high, we limit our analysis to identifying the immediate cause of the Rossby gyre oscillation and its effects on observables, and defer detailed analysis of the moist atmosphere feedbacks between clouds, convection, specific humidity, and the zonal wind, as well as further comparison to Earth analogues to future studies.

Our simulations of transit spectra show that the variable cloud cover caused by traveling Rossby waves could affect transit depths, though for our chosen planets the effect is too small to be observable with current instruments. \cite{may_water_2021} reported a transit depth variation due to cloud cover on the order of 10 ppm for their 10$^{-4}$ bars of CO$_2$ (with one bar of N$_2$ TRAPPIST-1e simulation with ExoCAM \citep{wolf_exocam_2022}, which is comparable to the THAI Hab 1 set-up \citep{sergeev_trappist-1_2022} and to our Control TRAP-1e experiment. THAI Part III \citep{fauchez_trappist_2021} found the standard deviation of the variation of the continuum level for Hab 1 to be 3 ppm for ExoCAM and 1 ppm for the UM, compared to our min-max difference of 1.26 ppm for for the 2.7 $\mu$m feature. \cite{song_asymmetry_2021} found the amplitude of temporal variability in their simulated transmission spectra, also for data generated by ExoCAM, to be on the order of 20 ppm. The slightly smaller degree of variation in our results compared to \cite{may_water_2021} and \cite{song_asymmetry_2021} is in line with the finding in \cite{sergeev_trappist-1_2022} and \cite{fauchez_trappist_2021} that ExoCAM displays the greatest degree of cloud variability out of the four models included in the comparison. Our quantitative findings agree with the results of these previous works and support the conclusion that atmospheric variability due to clouds will be below the noise floor of JWST.

From a qualitative perspective, the impact of Rossby waves on observations is highly sensitive to the location of both the Rossby gyres and the cloud deck. If clouds form on or extend to the planetary terminators, migrations by Rossby gyres could regularly clear this region for periods of time as long as the planet's transit. Based on our simulations, this scenario is most plausible on slowly-rotating planets where the gyre oscillation period is long and the atmospheric state at the planetary limb may persist for longer. A warmer planet with more cloud would show greater variability in transit depths as advection by the Rossby wave gyres results in long periods of heavy cloud cover and periods of entirely clear sky. Without prior knowledge of the cloud structure, cycle duration, and cycle phase, it is impossible to predict when clearing events might occur and to time observations to avoid flattened, featureless spectra due to clouds \citep{garcia_hstwfc3_2022, diamond-lowe_ground-based_2018, de_wit_combined_2016, kreidberg_clouds_2014, knutson_featureless_2014}. However, as the body of data from transit spectroscopy grows, atmospheric and climate variability should be considered when combining or interpreting data from different observing periods. In addition, as our theoretical understanding of climate variability on exoplanets improves, it may be beneficial to obtain data from consecutive transits instead of randomly chosen ones, as consecutive observations are more likely to represent a real atmospheric state rather than an averaged, composite one.

Traveling Rossby gyres could also affect the chemical composition of the planet's atmosphere. Several studies using chemistry-climate models have found that different chemical environments form on the dayside and nightside of tidally locked planets due to the presence or absence of photochemistry \citep{braam_lightning-induced_2022, yates_ozone_2020, chen_biosignature_2018}. In particular, the nightside gyres can build up high concentrations of species that are destroyed on the irradiated dayside \citep{ridgway_3d_2023}. In our simulations of TRAPPIST-1e, however, the nightside gyres frequently travel back and forth over the eastern terminator, exposing chemically enriched nightside air to radiation. This process may reduce chemical differences between the dayside and nightside, leading to a more homogeneous planetary climate.

The specific features of this atmospheric oscillation, including the period, the latitudes and longitudes of the Rossby gyres, their size, the distance they travel, and how much cloud (if any) they advect are dependent on model set-up, especially the cloud parameterization. The THAI project found significant differences in the cloud water paths predicted by the four models included in the intercomparison, with the UM in the middle of the pack \citep{sergeev_trappist-1_2022}. The location of the Rossby gyres also differs between models and between simulations with varying parameters. To date, only \cite{skinner_modons_2022} have studied the Rossby wave lifecycle in a tidally locked planet simulation. In their high-resolution, hot gas giant simulations, Rossby gyres (or ``modons'') fully circumnavigate the planet in the westward direction, periodically dissipate, and then reform and begin circulating again. We did not observe circumnavigation even in a matching high-resolution simulation performed with our control Proxima Centauri b model. Numerous factors such as the temperature structure of the atmosphere and the presence of clouds may influence the motion of Rossby waves in simulations of tidally locked planets. A better understanding of the sensitivity and evolution of these waves in atmospheric models of tidally locked planets is key to understanding climate variability and a fruitful avenue for future work.

\section{Conclusion}\label{sec:conclusion}
We describe a mechanism in the atmosphere of tidally locked terrestrial exoplanets in which feedbacks between clouds and incoming stellar radiation influence the dynamical state of the atmosphere, especially the zonal wavenumber-1 Rossby response to the thermal forcing, leading to alternating eastward and westward propagation of the Rossby gyres previously characterized as largely stationary. This proposed mechanism is as follows: 1) a decrease in substellar cloud cover reduces cloud albedo and increases the shortwave heating at the substellar surface of the planet; 2) the greater substellar atmospheric heating increases the peak-to-trough amplitude of the spatially periodic heating pattern, i.e. it increases the difference between dayside and nightside temperatures, and thereby enhances the Matsuno-Gill equatorial Rossby wave response, as shown in our spectral analysis of Rossby waves in the flow over time; 3) zonal mean wind speed also increases, shifting the Rossby wave response eastward of its equilibrium position and away from the substellar cloud region; 4) this results in decreased cloud cover at the eastern terminator. When substellar cloud cover increases again, the cycle ``runs in reverse'': shortwave heating and atmospheric temperature drop, the Matsuno-Gill response weakens, the zonal wind slows, and the Rossby gyres shift westward.

The oscillation in the location of the Rossby gyres can only affect the distribution of clouds if the path of the Rossby gyre migration intersects with the dayside cloud cover. In our simulations of Proxima Centauri b, this interaction results in periodic clear and cloudy days at the planet's eastern terminator, while in our simulations of TRAPPIST-1e, the Rossby gyres are located too far polewards to interact with the dayside clouds. Time series of synthetic spectra generated for a 300-day sample of the climate oscillation confirmed that the variation in cloud cover and atmospheric humidity associated with the feedback mechanism results in a time-varying transmission spectrum, but the magnitude of the variation in transit depths is too small to be detectable for our simulated planets. The mechanism is most likely to be observationally relevant on warm, slowly-rotating planets, as the long period of the Rossby gyre oscillation may clear the planetary terminators of cloud cover for extended stretches of time.

This study and our previous work in \cite{cohen_longitudinally_2022} identify physical mechanisms of variability which cause cycles in the planetary climate even in idealised exoplanet models without eccentricity, obliquity, or changes in the stellar spectrum. More complex environments on real planets are no doubt subject to additional sources of variability. As the body of exoplanet observations grows in the age of JWST and other upcoming telescopes, consideration of long-term climate variability and of weather on other planets can help interpret observations taken at different times, construct time series, and inform observing and data processing practices.

\section{Data availability}
A 200-day sample of model output data from each simulation presented in this study is available for public download at \url{https://doi.org/10.5281/zenodo.7752337}. The source code to generate the figures in this study is available at \url{https://github.com/maureenjcohen/cloudcode}.

%% Also note that the akcnowlodgment environment does not support long amounts of text. If you have a lot of people and institutions to acknowledge, do not use this command. Instead, create a new \section{Acknowledgments}.
\begin{acknowledgments}
We acknowledge the funding and support provided by the Edinburgh Earth, Ecology, and Environmental Doctoral Training Partnership and the Natural Environment Research Council [grant number NE/S007407/1]. We also kindly acknowledge our use of the Monsoon2 system, a collaborative facility supplied under the Joint Weather and Climate Research Programme, a strategic partnership between the Met Office and the Natural Environment Research Council. Our research was performed as part of the project `Using UKCA to investigate atmospheric composition on extra-solar planets' (ExoChem). This work was supported by a UKRI Future Leaders Fellowship [grant number MR/T040866/1], Science and Technology Facilities Council Consolidated Grant [ST/R000395/1] and the Leverhulme Trust through a research project grant [RPG-2020-82].'
\end{acknowledgments}


\vspace{5mm}


\software{We used the Iris Python package to manage and analyse model output data \citep{metoffice_iris_2010}. We used the windspharm Python library to perform Helmholtz decompositions to characterise the mean circulation \citep{dawson_windspharm_2016} and NASA's Planetary Spectrum Generator to simulate transit spectra \citep{villanueva_planetary_2018}.
          }




\bibliography{rwaves_bib}{}
\bibliographystyle{aasjournal}



\end{document}